# Conceptual design report on a charge breeder for HIE-ISOLDE

A. Shornikov[1],

E. Beebe[2], M. Breitenfeldt[1], J. Pitters[1], R. Mertzig[1], A. Pikin[2] and F. Wenander[1]

This work is part of CATHI Work Package 8: Radioactive Ion Beam Quality Improvement.
1: CERN, Switzerland
2: BNL, USA

## Abstract

In this conceptual design report the possible options for an upgrade of the REX/HIE-ISOLDE charge breeder are discussed. The performance requirements imposed by standard HIE-ISOLDE physics as well as injection into a possible future TSR@ISOLDE are discussed, and thereafter translated into machine parameters. Experimental results from tests of a high-current and high-density electron gun performed at Brookhaven National Laboratory are presented, and alternative gun designs are discussed. Finally, a cost estimate is given together with possible beneficiaries of the on-going R&D, and potential collaboration partners are identified.

The research leading to these results has received funding from the European Commission under the FP7-PEOPLE-2010-ITN project CATHI (Marie Curie Actions - ITN).
Grant agreement no PITN-GA-2010-264330.

# 1   Contents

## 2    About this report

This report is a 09/2016 revision of an internal reported prepared at the conclusion of the HIE-ISOLDE design study in 09/2014. The revised version includes latest experimental results of 07/2016 and gives a brief overview of other related activities such as the MEDeGUN project, which can have profound impact on the future charge breeder development. The requirements on the breeder for the TSR@ISOLDE proposal [4] are also included in this report.

## 3    Motivation

### 3.1    HIE-ISOLDE requirements

In this section we report on a part of the HIE-ISOLDE design study aiming to improve the radioactive ion beam quality by a high-performance charge breeder (CB). For our application we focus exclusively on a CB based on the Electron Beam Ion Source (EBIS) technology – an EBIS Charge Breeder (ECB). The main quality improvement requested by HIE-ISOLDE is an increase of the repetition rate and extension of the extraction pulse length from the ECB. While the superconducting part of the linac is CW, the normal conducting part is still pulsed with a maximum repetition rate of 50 Hz and an RF pulse length of 800 μs (present value, to be upgraded to 2 ms with HIE-ISOLDE). An extended extraction time ($t_{extr}$) will allow to use the entire RF time-window of the linac and therefore reduce the instantaneous counting rate ($CR$) on the detector $CR = N_{extr}/t_{extr}$, where $N_{extr}$ is the number of ions extracted from the charge breeder. Instantaneous CR above $10^6$ s$^{-1}$ [1] will cause pile-up at the MINIBALL detector. The instantaneous rate is further reduced by spreading the events over more pulses, realized by an increased repetition rate. The new ECB should retain capacity, acceptance and emittance close to those of the existing ECB, REXEBIS [2]. See a scheme of design-critical constraints in Figure 1.

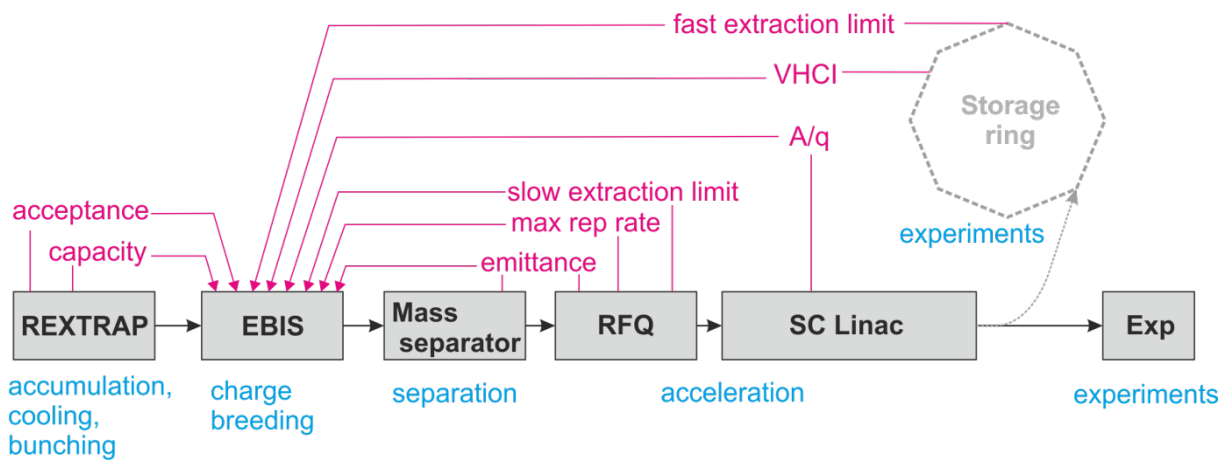

**Figure 1. REX-ISOLDE scheme and design constraints imposed on the ECB by other subsystems.**

### 3.2    TSR@ISOLDE requirements

Apart from the HIE-ISOLDE design study [3], there is a parallel independent proposal to install the collector storage ring TSR after the HIE-ISOLDE linac [4]. TSR@ISOLDE requests ions with higher charge state than REXEBIS can presently produce. The suggested program includes among other things [4]: bare ions from Z=30 through Z=70 for the astrophysical p-process capture, H and Li-like Cu, Sn, Tl for the study of atomic effects on nuclear half-lives and, Li or Na-like Lu, U, Th for di-electronic recombination on exotic ions. Apart from these most demanding experiments, TSR favors higher charge states for experiments with gas-jet collisions as the electron stripping probability from the stored ions is then reduced and the storage lifetime increased. The argument is also valid, although less pronounced, for in-ring decay experiments without a gas-jet, but with residual gas still present. A typical storage energy of the beam inside the ring is 10 MeV/u, dictated by either experimental conditions or lifetime considerations if a gas-jet is present. With a maximum rigidity of 1.57 Tm for TSR [4], only beams with $A/q$ < 3.5 can be stored [4]. Furthermore, higher charge states will also speed up the preparation of the ion beam after it has been injected into TSR as the electron cooling time, typically varying between 0.2 and 2 s, is given as t$_{cooling}$ ≈ $3A/q^2$.

As multi-turn ion injection into the ring will be used, the extraction pulse length out of the charge breeder has to be shorter than 30 μs in order to avoid ion losses during the injection process. The length of the injection window is inversely proportional to the transverse emittance from the breeder, thus longer extraction times can be accepted if the emittance is smaller and vice versa. The injection rate into the ring is defined by its operation mode, which can vary widely with different types of experiments. Two types of injection schemes, the first for reaction measurements using a gas-jet target inside the ring and the second for decay measurements without gas-jet, are given in Figure 2.

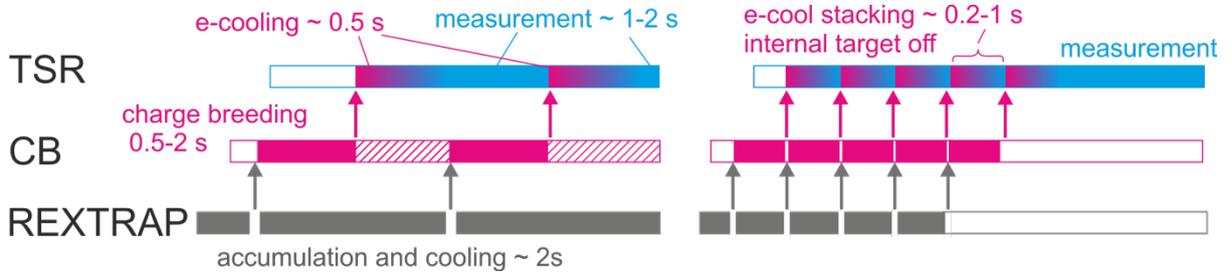

**Figure 2. Examples of injection schemes into the TSR.**

# 4    Technical design specification

## *4.1    Capacity and acceptance*

The capacity requirements of the new breeder should match the capacity of REXTRAP, where the primary 1+ ions are accumulated and bunched. The experimental capacity and emittance of REXTRAP measured with potassium ions using neon buffer gas are reported in ref. [5]. As shown in Figure 3A even though some $10^9$ ions can be injected into the REXTRAP, only about $10^8$ can be extracted per bunch due to losses growing with the number of stored ions [5]. The number of stored ions also affects the emittance of the extracted beam (see Figure 3B). For high intensities, the cooling becomes inefficient and the emittance of the REXTRAP is close to the emittance of the ion beam extracted from the primary target-ion-source.

Such large transverse ion-acceptance is impossible to achieve with a high compression electron beam. Even a low compression beam as REXEBIS at typical operation conditions (current 0.25 A, current density 125 A/cm², electron energy 4 keV, magnetic field $B$=2 T), has only an acceptance of α = 11.5 microns for ions injected at $U_{ext}$ = 30 keV according to:

$$\alpha = \frac{r_{beam}}{\sqrt{2U_{ext}}}\left[Br_{beam}\sqrt{\frac{q}{M}} + \sqrt{\frac{qB^2r_{beam}^2}{4M} + \frac{\rho_l}{2\pi\varepsilon_0}}\right]$$

where $\rho_l$, $r_{beam}$ and $M$ represent linear charge density, electron beam radius and the ion mass.

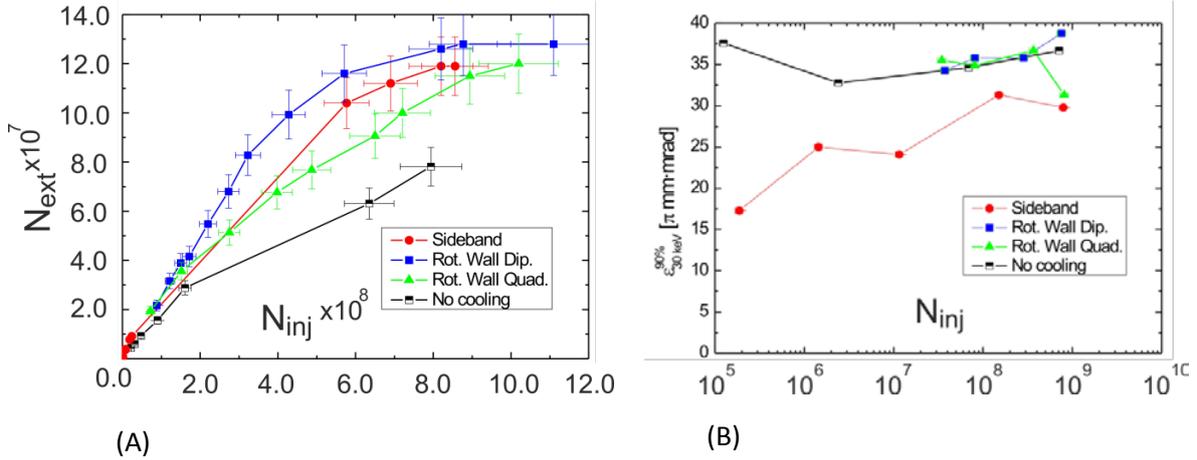

**Figure 3. (A) Number of ions extracted from REXTRAP as a function of the number of injected ions. (B) 90% emittance of the REXTRAP as a function of the number of injected ions [5].**

The pulsed ion injection, however, relaxes the injection condition and accepts trapping of ions, even if the ions are injected only with partial overlap with the electron beam. A realistic goal for the ion acceptance of the upgraded high compression electron beam is a value similar to the acceptance obtained with the REXEBIS immersed low-current beam.

For the capacity estimation we consider the worst-case scenario of 89$^+$ ions, confined by a 150 keV electron beam with a neutralization factor of $f$ = 0.1. In the present design version, we consider a trap length $L_{trap}$ of 1 m, compared to earlier suggested 0.7 m. The capacity in elementary charges can be calculated as:

$$N = f I_e L_{trap} / e v_e$$

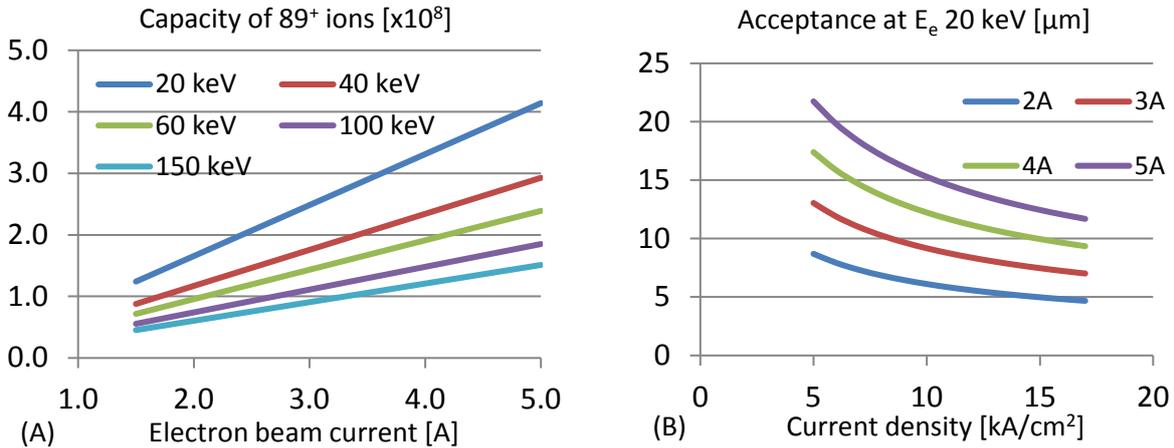

**Figure 4. (A) Capacity of an EBIS for 89$^+$ ions (·10$^8$) as a function of electron beam current for various electron beam energies. (B) Transverse ion beam acceptance of an EBIS (μm) as a function of the electron beam current density for various electron beam currents at 20 keV energy assuming 150 atomic mass unit ions.**

where $v_e$ is the electron velocity, $I_e$ is the electron current and $e$ is the elementary charge. As shown in Figure 4A an $I_e$ of 3.5 A will be sufficient to achieve the requested capacity even with an electron beam energy of 150 keV. Increasing the length further is technically challenging and the space limitations on the elevated platform for the charge breeder in the ISOLDE hall prohibit a total length of the breeder exceeding 4 m. A 3.5 A beam also fulfils the acceptance requirements with the requested current density of about 10 kA/cm² (see Figure 4B).

## 4.2   Ionization factor and repetition rate

The maximum repetition rate at which a specific ion can be provided to the users depends on the speed of the charge breeding process. For the assumption of successive ionization of all charge states from the initial one to the final, we can introduce the ionization factor as the product of electron current density $J_e$ and the time $t$ an ion is exposed to the electron beam. This is a universal criterion for achieving a certain charge state, according to which the only difference between low-density and high-density electron beams is the time it takes to produce a certain charge state. In reality, however, it is not true due to competing recombination processes. Out of two main recombination mechanism counteracting the ionization, radiative recombination with beam electrons and charge exchange with neutrals, only the radiative recombination (RR) is proportional to the ionization factor. The presence of charge exchange prevents breeding-time scaling for various current densities. The highest charge states can only be produced with high-density electron beams. We will discuss it more in the dedicated subsection "Charge exchange, vacuum and beam neutralization". In this subsection, we consider a simplified case when the charge exchange process is neglected.

We consider two cases: the HIE-ISOLDE and the TSR@ISOLDE scenarios. The first is defined by the wish to perform fast breeding of e. g. Ba ions to an A/q < 4.5 (imposed by the linac) within 10 ms; the second by production of Li-like U ions within 1 s. The current density $J_e$ required to produce these charge states can be calculated as follows:

$$J_e = \frac{e}{t_q} \sum_{j=1}^{q} \frac{1}{\sigma_j^{II}(E_e)}$$

where $e$ is the electron charge, $\sigma_j^{II}(E_e)$ is the electron impact ionization cross-section of charge state $j$ by electrons with the energy $E_e$. Using Lotz's formula [6] for $\sigma_j^{II}$ we obtain results presented in Figure 5. By coincidence, both cases require about the same $10^4$ A/cm².

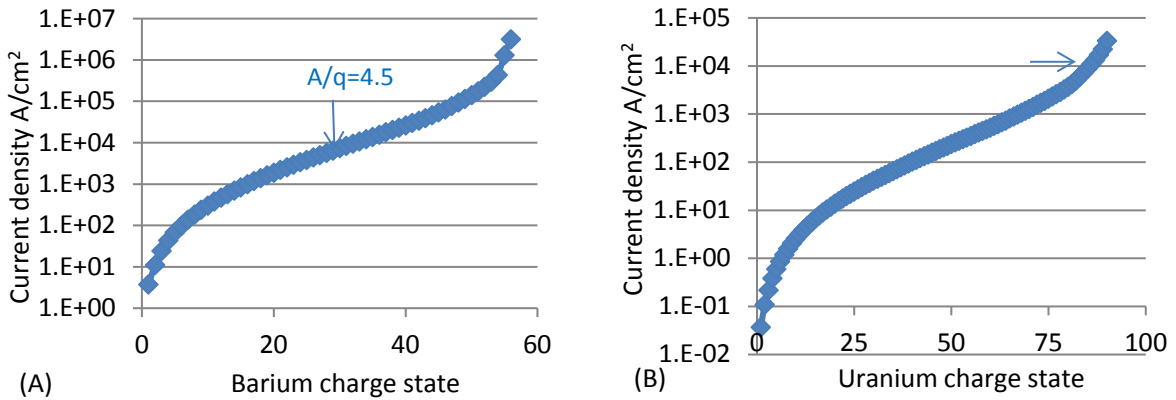

**Figure 5. (A) Current density required to produce Ba ions with certain charge state within 10 ms. (B) Current density required to produce U ions with a certain charge state within 1 s.**

### 4.3 Energies for HIE-ISOLDE and TSR@ISOLDE scenarios

The ionization energy of the desired charge state determines the lowest limit of the electron energy of the ECB. Operating at the lower limit is inefficient, as the radiative recombination process counteracts the ionization process. Both processes scale with ionization factor. The only way to shift the balance towards ionization is to increase the electron energy such that the cross-section of ionization is increased while the recombination is suppressed. If we consider the highest charge states of an ion species in equilibrium between electron impact ionization and radiative recombination, we can reconstruct the charge state distribution as a function of the electron energy as follows:

$$N_q = N_{q-1} \frac{\sigma_{q-1}^{II}(E_e)}{\sigma_q^{RR}(E_e)}$$

where $N_q$, $N_{q-1}$ are charge state abundances, $\sigma_j^{II}(E_e)$ and $\sigma_j^{RR}(E_e)$ are energy-dependent impact ionization (II) and radiative recombination (RR) cross-sections. Using Lotz formula [6] for $\sigma_j^{II}$ and Stoehlker [7] for $\sigma_j^{RR}$ we obtain the charge state distributions of the highest charge states for Ba and U (see Figure 6).

The abundance of a charge state has a maximum at an electron energy about 2.7 times higher than the ionization energy of that charge state owing to a maximum in the ionization cross-section according to Lotz's formula [8]. To reach some of the requested species with 30-40% abundance electron energies up to 150 keV will be required. It should be noted that Lotz formula predicts cross-sections near their maximum with +40/-30% [6] uncertainty. The Lotz's formula was derived for low charge states and there are controversies about its prediction power for high charge states. Some researchers state that Lotz overestimates the cross-section for high charge states [9], while in experiments by Marrs et al. [10] it was shown that for He-like and H-like U the Lotz formula underestimates the cross-section by a factor of 2. The same uncertainties apply to calculated ion heating, as it includes ionization time, inversely proportional to the sum of the ionization cross-sections.

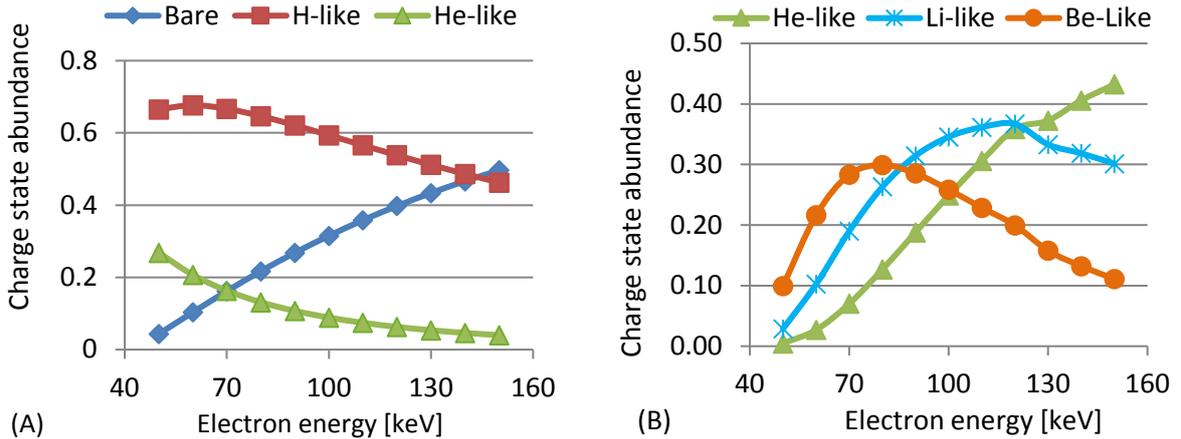

**Figure 6. Charge state abundances in dynamic equilibrium between impact ionization and radiative recombination as a function of the ionizing electron energy. (A) Ionization of Ba. (B) Ionization of U.**

### 4.4 Charge exchange, vacuum and beam neutralization

As mentioned earlier, charge exchange (CX) with neutral atoms is a recombination mechanism, which does not scale with the ionization factor. Thus, the prominence of CX is higher for low current densities. The CX cross-section grows fast with the charge state and the process is comparable to ionization or radiative recombination events at practical beam and vacuum parameters. Other features of the CX process is rather poor fundamental data on the cross-sections, approximate formulae derived from fittings, significant discrepancies (about 50%) between various data sources, large error bars of ±80% in many cases and lack of primary data charge states above q = 8-10 [11], [12], [13].

Even using the best available data verified up to charge states 43+ [14], [15], we need to allow for safety margins. Let us consider a 150 keV electron beam with $10^4$ A/cm² current density ionizing U ions. For any given charge state we can find at which vacuum pressure the probability of charge exchange recombination is equal to the probability of the electron impact ionization. Similarly, we can find a pressure at which the probability of radiative recombination is equal to the probability of charge exchange. Both equilibrium curves are shown in Figure 7. For higher charge states the balance of ionization and charge exchange is established at high $10^{-10}$ mbar at this particular current density. Close to that region charge exchange also approaches the efficiency of radiative recombination and at practical pressures may become dominant if the current density will be lowered (thus the assumptions used for plots shown in Figure 6 of ionization and RR equilibrium will be violated). Keeping in mind the limitations of primary charge exchange data, we should specify the desired base pressure in the ionization region to $10^{-11}$ mbar. While being a significant constraint this seems feasible as pressures in that range have been demonstrated by REXEBIS under usual operational conditions.

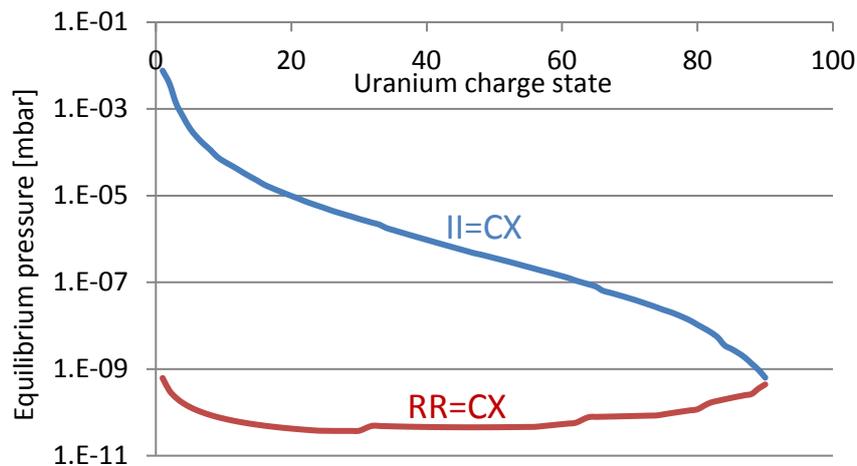

**Figure 7. Equilibrium pressures. At the equilibrium pressure the probability of charge exchange for a certain U charge state is equal to the probability of radiative recombination (red line) or impact ionization (blue line). For impact ionization and radiative recombination we assume an electron energy of 150 keV and an electron current density of $10^4$ A/cm².**

Another practical question is whether the ions created from ionization of the residual gas will generate significant space charge that affects the neutralization of the negative electron beam charge. We consider two cases, short 10 ms cycles with 20 keV electrons (HIE-ISOLDE beams) and 1 s cycles with 150 keV electrons (TSR@ISOLDE beams). In both cases we assume a 3.5 A current and $10^{-10}$ mbar residual pressure giving a gas density $n$ = 2.4·$10^6$ particles per cubic centimeter. We furthermore assume that the residual atmosphere is composed entirely of hydrogen. As our working electron energies are much higher than the ionization energy of hydrogen ($E_i$ = 13.6 eV) we will use the Bethe asymptotic approximation of the total ionization cross-section $\sigma_{TICS}$ :

$$\sigma_{TICS} = 4\pi a_0^2 \left[ A \frac{\ln(u)}{u} + \frac{B}{u} + \frac{C}{u^2} \right]$$

where $a_0$ is the Bohr radius of $5.29 \times 10^{-11}$ m, $u = E_e/E_i$, and A, B and C are Bethe coefficients for hydrogen equal to 0.2834, 1.2566 and -2.63 respectively [16]. This yields $\sigma_{TICS}$ of $7.94 \cdot 10^{-19}$ and $1.24 \cdot 10^{-19}$ cm$^2$ for 20 and 150 keV respectively. Using these cross-sections we can calculate the production of ions as a current:

$$I_{res} = n\sigma_{TICS} I_e L_{trap}$$

which gives 0.67 and 0.1 nA for 20 and 150 keV respectively. Assuming all ions are trapped throughout the breeding period $\tau$, the accumulated charges will be 6.7 pC and 100 pC, corresponding to an additional neutralization of $1.6 \cdot 10^{-4}$ and $6.5 \cdot 10^{-3}$ respectively. This shows that partly owing to the very high electron energies (thus inefficient low charge-state ionization), the beam compensation is not a limiting factor for this ECB, in contrast to early EBISes like the Saclay CRYEBIS [17].

For the design purposes we conclude that a vacuum level of lower $10^{-10}$ mbar will be sufficient to avoid the effects of charge exchange and beam neutralization for the ECB.

### 4.5  Emittance, ion heating and cooling

The emittance of the beam produced by the ECB is defined by the breeding process and the geometrical beam properties. There are several processes that will introduce energy spread to the beam. First, due to random radial positions at ionization from charge state q to q+1, the ion cloud will gain energy [18]. Assuming uniform electron current density across the electron beam, and a harmonic oscillator ion motion in the radial potential, then - according to ref. [18] - the temperature change per ionization step is $\Delta kT = V_0/5$ where $V_0$ is the radial space charge potential from the beam axis to the edge. For a 3.5 A and 150 keV electron beam $V_0$ is equal to 100 V, so a temperature of 20 eV/q will be achieved. Compared to the shallow (~10 eV) potentials in high energy and low current SuperEBITs, the ionization heating in the elaborated ECB is large, however, since it scales with the space-charge potential depth the ion confinement is of little concern in both cases. On the other hand, the ions extracted from an EBIS with a deeper space charge potential will have larger emittance due to this process.

The second energy transfer phenomenon is individual binary scattering of electrons on ions, which transfers energy to the ions (also called Landau-Spitzer heating). This process is proportional to the ionization factor and therefore cannot be suppressed. We can calculate the energy transfer [19] as follows:

$$\Delta kT = \frac{10^{-18}}{AE_e} \sum_{i}^{i=max} \frac{i^2}{\sigma_i^{II}}$$

where $A$ is mass in atomic mass units, $E_e$ in eV and $\sigma_i^{II}$ in cm$^2$. For uranium it gives a curve as shown in Figure 8A.

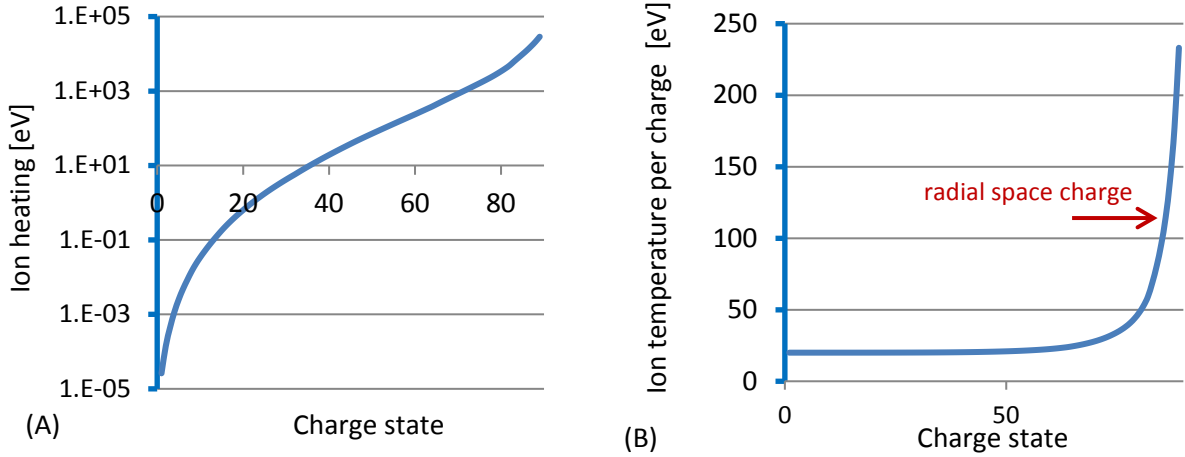

**Figure 8. (A) Total Landau-Spitzer heating of uranium ions as function of reached charge state. Temperature per charge for uranium ions as function of charge state when accounting for Landau-Spitzer and ionization heating.**

The mean temperature rise per charge is about $\Delta E$ = 240 eV for Li-like U. Adding ionization heating $\Delta kT = eV_0/5 = 20$ eV we obtain 260 eV/q, which exceeds $V_0$ by 160 eV/q. Thus, to preserve the ions and achieve a Li-like state, evaporation of light ions from the trap is needed to cool the heavy ions. This technique was successfully used for bare U production in an EBIT [20] and other species in EBIS devices like Dioné (Saclay) [21], and in the EBIS derivative Electron String Ion Source (ESIS) [22]. Light ions may come either from the residual gas or be deliberately injected as a gas-jet. The required evaporation rate of light ions is estimated with the following argument. Each light gas atom will initially be ionized at a random position, with an average potential energy of $0.5 \cdot V_0$. The remaining $0.5 \cdot V_0$ will be taken by evaporation. Therefore, to remove 160 eV/q per $U^{89+}$ ion we need 284 protons. The required rate means that only $2.2 \cdot 10^6$ $U^{89+}$ can be cooled if the only source of the light ions is the residual gas with the ionization rate calculated earlier at $10^{-10}$ mbar pressure. If the trap is filled to the design capacity additional gas injection is required. The optimal way for such injection is to use a gas-jet [10]. This allows keeping a low base pressure while introducing cooling ions. When choosing the cooling gas one should also keep in mind that besides taking less energy per ion, lighter ions have less efficient energy transfer mechanism from heavy ions as the heat transfer depends on $Z_c^2 Z_h^2$, where $Z_c$ and $Z_h$ are the charge states of cooling and heavy ions respectively. So an ideal gas would be nitrogen or neon and not protons.

If the ions are cooled to stay within the electron beam their emittance can be calculated as $r_{beam}\sqrt{\Delta E/U_{ext}}$ where $\Delta E$ is the energy spread, $U_{ext}$ is the extraction energy, and $r_{beam}$ is the electron beam radius. The ion energy spread is defined by the radial potential as long as the axial trapping barriers are higher than the radial potential, which they are in order for the EBIS to fully utilize the space charge capacity. For charge breeding of $U^{89+}$ assuming a 150 keV, 2.7 A electron beam, the space charge potential difference between the beam center and its edge is about 100 V. Cooled ions have an energy spread of $\Delta E \approx 0.1\text{-}0.4\ qV_0$ [18], but we conservatively assume $\Delta E \approx qV_0 = 100\ eV$. For injection into the REX-ISOLDE RFQ an energy of 5 keV/u is requested, corresponding to $U_{ext}$ = 13.3 kV for $U^{89+}$. Assuming a 0.1 mm electron beam, we obtain an emittance of 8.7 μm (0.029 μm normalized).

This is less than the 0.08 µm normalized acceptance of the downstream A/q-separator and RFQ [23]; see ion beam tracing plots and acceptance ellipses in Figure 9. For breeding of lower charge states at lower beam energies of around 50 keV, the space charge potential will double and the emittance grows by $\sqrt{2}$ in the absence of evaporative cooling. Still, sufficient safety margin of the acceptance will allow to transfer the beam.

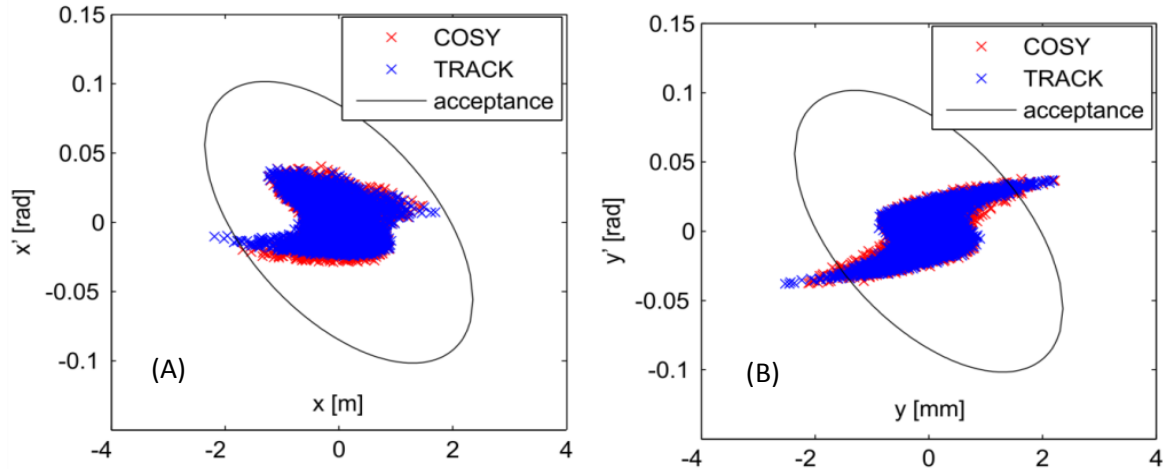

**Figure 9. Phase space of an ion beam traced through the A/q separator and RFQ. Tracking simulations in vertical (A) and horizontal (B) planes using COSY and TRACK codes. The ion beam has a normalized emittance of 0.08 µm and an energy spread of 1% [23].**

Optics simulation [23] have demonstrated that an energy spread in excess of 1% will cause beam losses. For $U^{89+}$ an energy spread of about 1.8% can be expected due to Landau-Spitzer and ionization heating if the beam is not cooled. Therefore, for Very Highly Charged Ion (VHCI) applications ion-ion cooling will be required not only to avoid beam losses, but also to improve the extracted beam properties. Reduced ion velocities from the ion-ion cooling will also suppress charge exchange processes with neutrals inside the breeding region, especially important for VHCI.

We see that an ECB can provide an ion beam quality suitable for further transport. In this section we considered only individual collisions as heat transfer phenomena. These phenomena cannot be avoided and the heat introduced by these processes have to be removed by evaporating ions.

In an EBIS-like device there is also a possibility of developing collective effects in multicomponent electron-ion plasma. The collective effects may occur only under certain conditions defined by the ECB design. The relevance of such risks for the upgraded ECB is analyzed below.

### 4.6    Plasma instabilities

An EBIS is a system at risk of several plasma instabilities [24], for instance two stream instability (TSI). The TSI may occur when one charge distribution moves with a certain velocity relative to another distribution at rest. This is exactly the case when the electron beam travels through the trapped ion cloud. If the TSI occurs, the energy of the electron beam will be transferred to the ion cloud by means of collective plasma oscillations. The possibility of plasma instabilities is one of the main motivations to continue full scale experiments on a test stand after electron beam parameters close to design specification have been achieved.

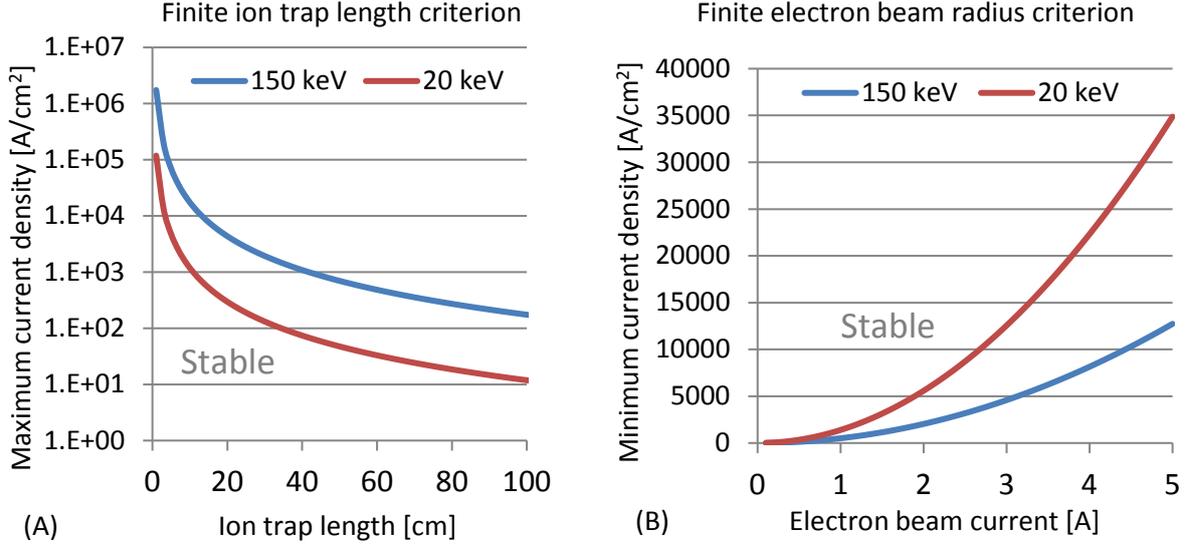

**Figure 10. TSI stabilization criteria based on finite ion trap length (A) and finite electron beam radius (B). The stable regions are indicated, above the curves in (A) and below in (B).**

For EBIS devices the potential danger of TSI was intensively studied in the past. The plasma modes excited by an electron beam using solid rotator approximation were considered in ref. [25]. The growth of the instability was found convective, i.e. the instability grows in amplitude while moving with the group velocity along the axial direction, as compared to an absolute instability, which grows at a specific location (see ref. [26]). The convective instability is unlikely to be disruptive for the electron beam transport, but it may cause ion heating and related losses and emittance growth. Compared to infinite plasma systems, the finite size of an EBIS may have stabilizing effects on the instability growth. Two criteria were suggested. The first is based on the finite length of the ion cloud. If the length is short enough, a perturbation will leave the system before growing significantly. The second takes into account the finite beam size, which may be smaller than the Debye screening length in the plasma. Therefore a set of constraints on the EBIT geometry were suggested, and LLNL-like super-EBITs were built very short with 2-5 cm trapping regions. Longer traps are believed to be prone to TSI. Successfully operational EBISes such as RHIC-EBIS, CARIBU or REXEBIS are unstable according to this criterion, however, the criterion is not applicable to immersed flow electron beams because it is derived for a solid rotator model of the electron beam (like a Brillouin beam).

Let us consider the stabilization criteria for the HEC² ECB [27]. The instability growth rate is given by:

$$\gamma = \frac{\sqrt{3}}{2}\left[\frac{\omega_{pe}\omega_{pi}^2}{2}\right]^{1/3} = 5.6 \times 10^{-2}\left[f\varphi\frac{q}{A}\right]^{1/3}\omega_{pe} = 1.8 \times 10^{8}\left[f\varphi\frac{q}{A}\right]^{1/3}\frac{j_e \ [A/cm^2]}{E_e^{1/4}\ [keV]}$$

where $\omega_{pe}$ and $\omega_{pi}$ are electron and ion plasma frequencies, $f = 0.1$ is the space charge neutralization factor and $\varphi = 1$ is the electron and ion beam overlap factor, A is the ion mass number and q is the charge. The length criterion states that the trapping region length $L$ should be less than:

$$L < \frac{3\,v_{phase}}{\gamma} \approx \frac{v_e}{\gamma}$$

where $v_{phase}$ and $v_e$ are the phase velocity of the instability and the electron velocity, respectively. This can be translated into a stability criterion for the trap length:

$$L \; [cm] \; < \; \sqrt{106 \frac{E_e^{3/2} \; [keV]}{\left[ f \varphi \frac{q}{A} \right]^{2/3} j_e \; [A/cm^2]}}$$

Hereafter we will consider the two cases relevant to the HIE-ISOLDE and TSR@ISOLDE applications (both with $10^4$ A/cm$^2$ and 3.5 A, and with 20 and 150 keV electron energies, respectively). For both cases HEC$^2$ ECB is unstable according to the length criterion. Only at a maximum length of about 3-12 cm the criterion will be fulfilled, as seen in Figure 10. For the second criterion we need to compare the electron beam size ($r_b$) with the Debye screening radius. The Debye radius is given by:

$$r_D = \sqrt{\frac{T_e}{4 \pi n_e e^2}}$$

where $T_e$ is the electron temperature in energy units, $n_e$ the electron number density and $e$ the elementary charge. Assuming a cathode radius $r_c$ of 10 mm and an electron emission temperature of 0.1 eV, corresponding to the bulk temperature of the cathode, the beam compression increases the temperature as:

$$T_e[keV] = 10^{-4} \times \frac{\pi r_c^2 j_e}{I_e} \; [cm, A/cm^2, A]$$

Hence the stability criterion is given by:

$$\frac{r_b}{r_D} = 1.4 \times \left( \frac{I_e^2 \; [A]}{10^{-4} \pi r_c^2 j_e \sqrt{E_e \; [keV]}} \right)^{1/2} < 1 \qquad \rightarrow \qquad j_e > \frac{1.96 I_e^2}{10^{-4} \pi r_c^2 \sqrt{E_e}}$$

The result is shown in Figure 11 for a 20 keV electron beam, as injection into 150 keV is not foreseen due to the low transverse acceptance. In Figure 11A the original HEC$^2$ configuration is shown. The blue curve indicates the current density limit according to Hermann formula, and the green line shows the relation between the current and current density such that the desired acceptance is obtained. For solid curves the viable parameter space is below the curves. The dashed red curve is the TSI stability criteria and the dashed black line at 10 kA/cm$^2$ is the threshold required to maintain the pulse repetition rate. For dashed curves the viable parameter space is above the curves. One can see that with the original requirements there is no overlap between the regions where all criteria are met. This is because the acceptance and TSI-stability put opposite requirements on the electron beam radius. With some modifications a narrow working region can be found, as shown in Figure 11B, where all limits are met. The following parameters are then assumed (original values in brackets): main magnet field 6 T (5 T), cathode radius 8 mm (10 mm), and acceptance 7.5 μm (11 μm) for ions with 150 atomic mass units and 30 kV extraction energy from REXTRAP (corresponds to 4.8 nm (7.2 nm) normalized acceptance). The combination is not unique. In all cases the residual B-field on the cathode was assumed to be 0.3 mT. A significant increase of the field makes it impossible to realize a suitable working region. In this analysis we have assumed that the electron temperature is only changing due to beam compression, with no additional spread. For the acceptance calculation we have neglected the possibility of a low compression region where a larger electron beam radius will promote the ion capture. Such a low compression section, similar to MSU ReA EBIT/S, can be deliberately introduced. While not stable relative to TSI due to the large electron beam radius, such a section may be used only to capture 1⁺ ions, and in this case the total neutralization will be about 100 times less than assumed in Figure 10. With reduced neutralization the trap length criteria will be relaxed allowing some 20 cm stable region with a current density of a few thousand A/cm².

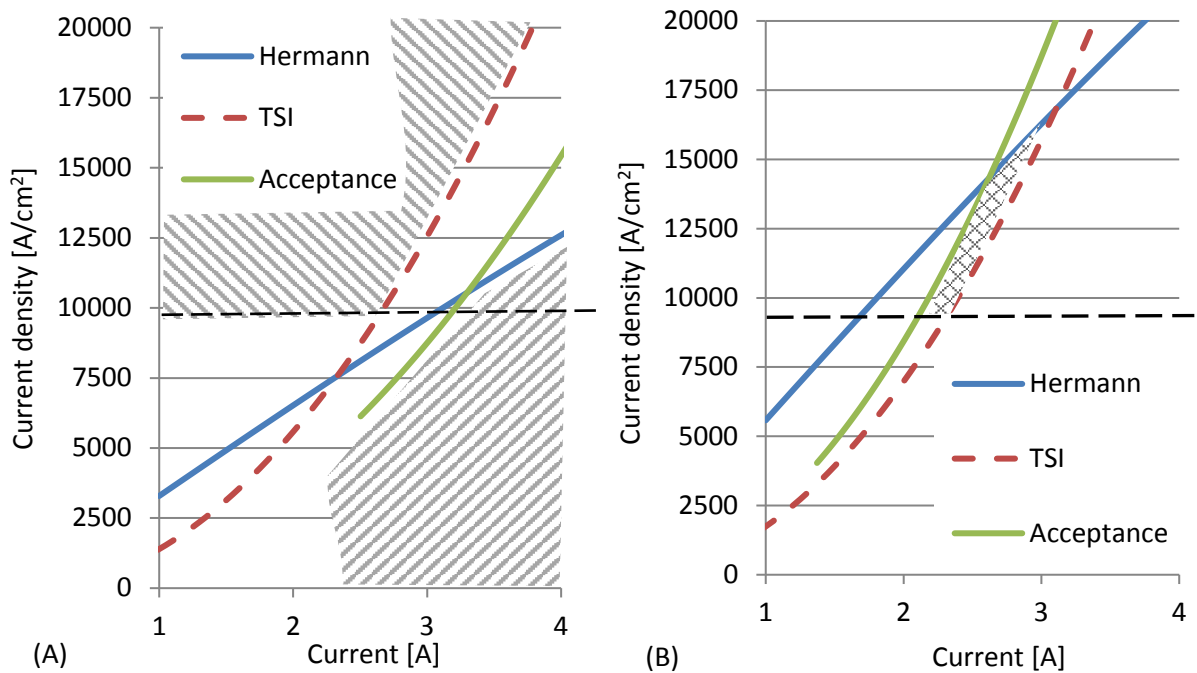

**Figure 11. (A) Present configuration. (B) Possible adjusted configuration. See main text for explanations.**

### 4.7   Technical design summary

The technical numbers from previous subsections are summarized in Table 1.

**Table 1. Technical design summary of the key parameters of an upgraded ECB**

| Parameter | Upgrade | Comments | REXEBIS |
|---|---|---|---|
| Electron energy [keV] | 20-150 | Ramped down for 1⁺ injection | 5 |
| Electron current [A] | 2.7 | | 0.25 |
| Electron current density [A/cm²] | >1.0·10⁴ | | 100 |
| Transverse acceptance [μm] | 7.5 | 90% non-normalized at 30 keV, pulsed injection for 150 amu mass | 11.5 |
| Ion-ion cooling needed | Yes | For VHCI only | No |
| Extraction emittance [μm] | <0.03 | 1 σ, normalized at 5 keV/u | |
| Extraction time [μs] | <30 to 2000 | Variable, application specific | >50 |
| Extracted energy spread | <1% | 1 σ, at 5 keV/u | |
| Vacuum [mbar] | <10⁻¹⁰ | Suppressing CX for VHCI | Low 10⁻¹¹ |

It should be noted that neither of the design parameters is isolated and they are mutually influencing each other. Typical time structures for the ECB cycle for HIE-ISOLDE and TSR@ISOLDE operation is given in Figure 12.

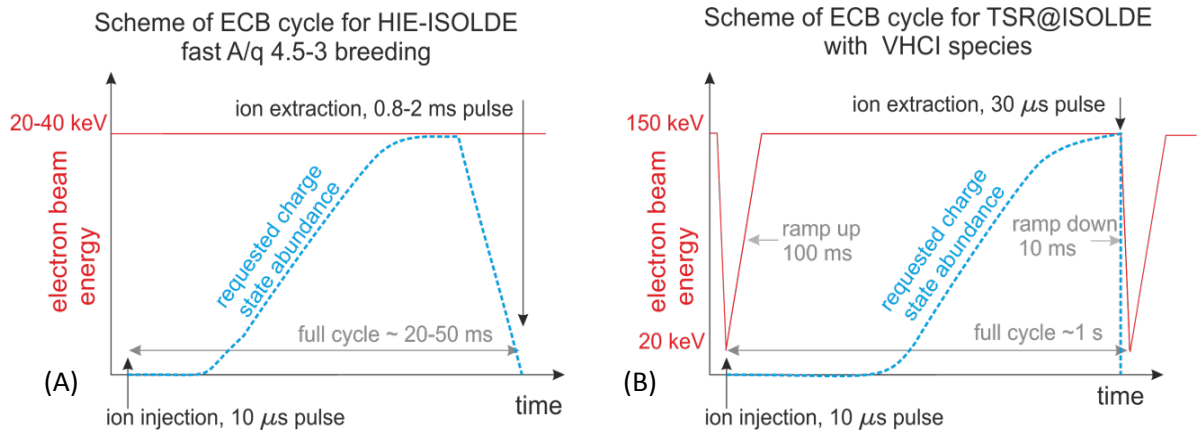

**Figure 12 . Typical time structures for the charge breeder operated for HIE-ISOLDE (A) and TSR@ISOLDE (B).**

## 5    Achieved results

The key to the upgraded ECB is a suitable electron beam. Such high performance beam requires switching from reliable and well-studied immersed beam technology (used at all contemporary operational high-capacity EBISes including REXEBIS, RHIC EBIS, CARIBU) to a Brillouin-like beam using combined electrostatic and magnetic beam compression to achieve the desired current density. Even though Brillouin-like beams with similar current density have been produced in EBITs, they were lower in current by a factor of 15 to 30 (see relative comparison of EBIS and EBIT devices in Figure 13). During the design study, we have been in close contact and established productive co-operations with EBIT/S groups worldwide, including groups at BNL, MSU-NSCL, MPIK, JINR, LLNL, ANL, MSL, University of Fudan (UF), University of Electronic Science and Technology of China, Chengdu (UESTC), Frankfurt University (FU), and Tokyo University of Electrical Communications (TUEC).

Several potential options were considered and the most promising of them were studied. Some options were marked as risky. Amongst the available options the most promising one was to team up with the BNL Advanced Ion Sources group in an attempt to build and test a prototype of a high performance electron gun, HEC[2].

We would like to stress that unlike many other guns for EBIT/S, the HEC[2] gun was tested on an actual EBIS from the very beginning. Compared to testing on low-field magnetic test stands it gave us the advantage of the full magnetic field. In some cases beams of compromised quality can be propagated in low-field test devices, but operation in full field is impossible. With our experimental approach we identify such problems at an early stage. As such the commissioning may be longer and more complicated, but the resulting solution should be fully functional. The obvious drawback is the necessity of the full scale test bench. Our project took advantage of the BNL TestEBIS – the only device worldwide suitable as a full-scale test bench for HEC[2]. The efforts and costs of shipping equipment and sending personnel were negligible compared to setting up an own equivalent device. But as a consequence, planning of experiments, scope and timing are strongly influenced by the availability of BNL personnel and partly by the machine legacy, such as power supplies or optics.

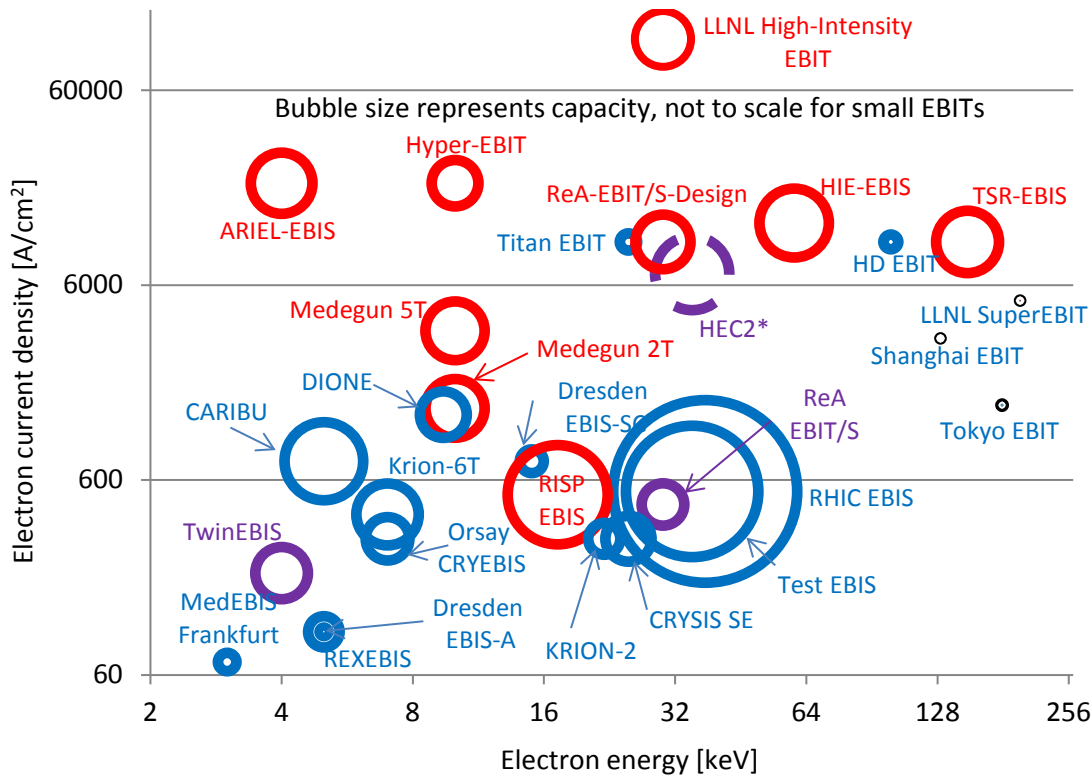

**Figure 13 .** Comparison of Electron Beam Ion Sources, Electron Beam Ion Traps and Electron String Ion Sources based on their capacity electron beam energy and current density (note log-log scale). Red circles denote project and design values, blue circles denote operational devices, purple circles denote devices in test operation and a dashed circle indicates that the current density value is based on simulations. The device performances are given according to ARIEL-EBIS [28], CARIBU [29], CRYSIS SE [30], DIONE [21], Dresden EBIS-A [31], Dresden EBIS-SC [32], HD EBIT [33], HEC2 [34], HIE-EBIS [27], Hyper-EBIT [35], KRION-2 [36], KRION-6T [37], LLNL High-intensity EBIT [38], LLNL SuperEBIT [10], MedEBIS Frankfurt [39], MEDeGUN 2T [40], MEDeGUN 5T [40], Orsay CRYEBIS [41], ReA-EBIT/S [42], ReA-EBIT/S Design [43], REXEBIS [44], RHIC EBIS [45], RISP EBIS [46], Shanghai EBIT [47], Test EBIS [48], Titan EBIT [49], Tokyo EBIT [50], TSR-EBIS [51], TwinEBIS [52].

## 5.1   Experimental performance of a test gun at BNL

In the early stage of the project we found out that a HEC²-like gun project had been initiated for future upgrades at BNL [53]. In a joint effort a high compression electron-gun of BNL design, was built at CERN [51] (see Figure 14). The gun underwent mechanical test at CERN and was shipped to BNL, accompanied by two CERN members.

After two installation and commissioning campaigns, each lasting 1.5 months with heavy involvement of the BNL technical team, the gun was operated for the first time. The gun was assembled and mounted onto the BNL TestEBIS (see Figure 15). The mounted gun underwent low temperature pre-baking, high voltage training, and recalibration of the cathode operation regime for an enclosed environment. In parallel the entire EBIS was conditioned for operation and a set of diagnostics tools was prepared. At the end of the second campaign the gun was operated for 6 days reaching a 1.5 A electron current [54], see Figure 16.

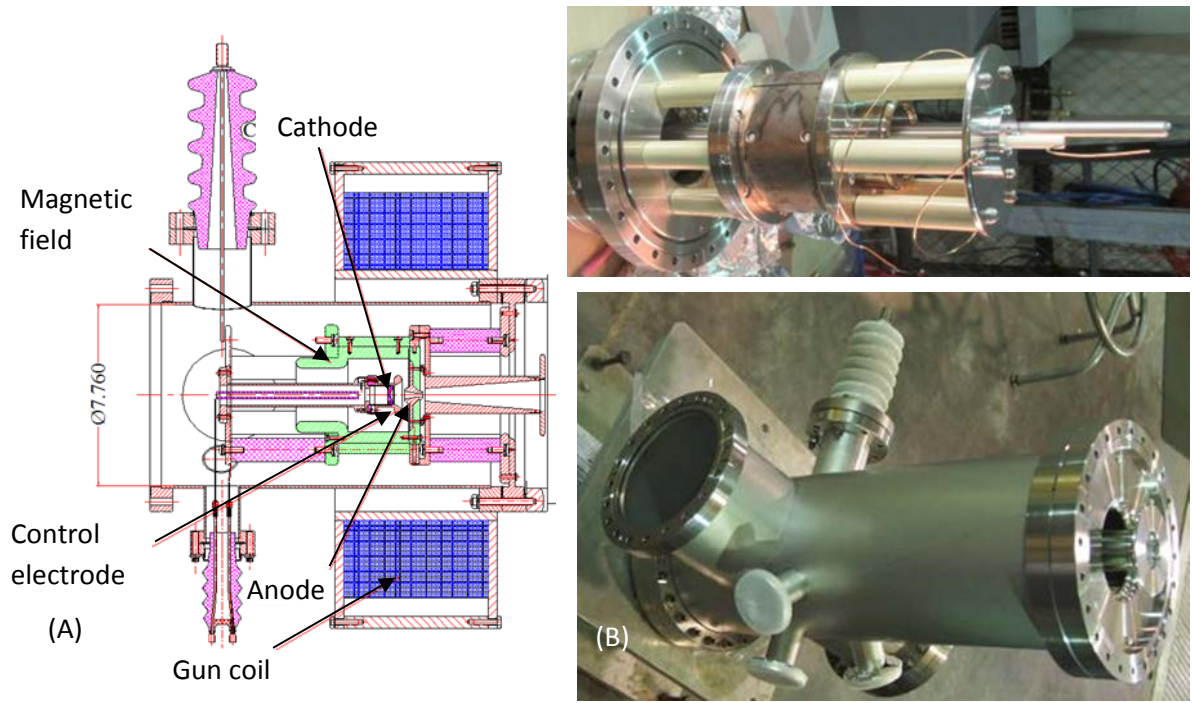

**Figure 14. Electron gun drawing. Gun assembled (B top) and installed in the gun chamber (B bottom).**

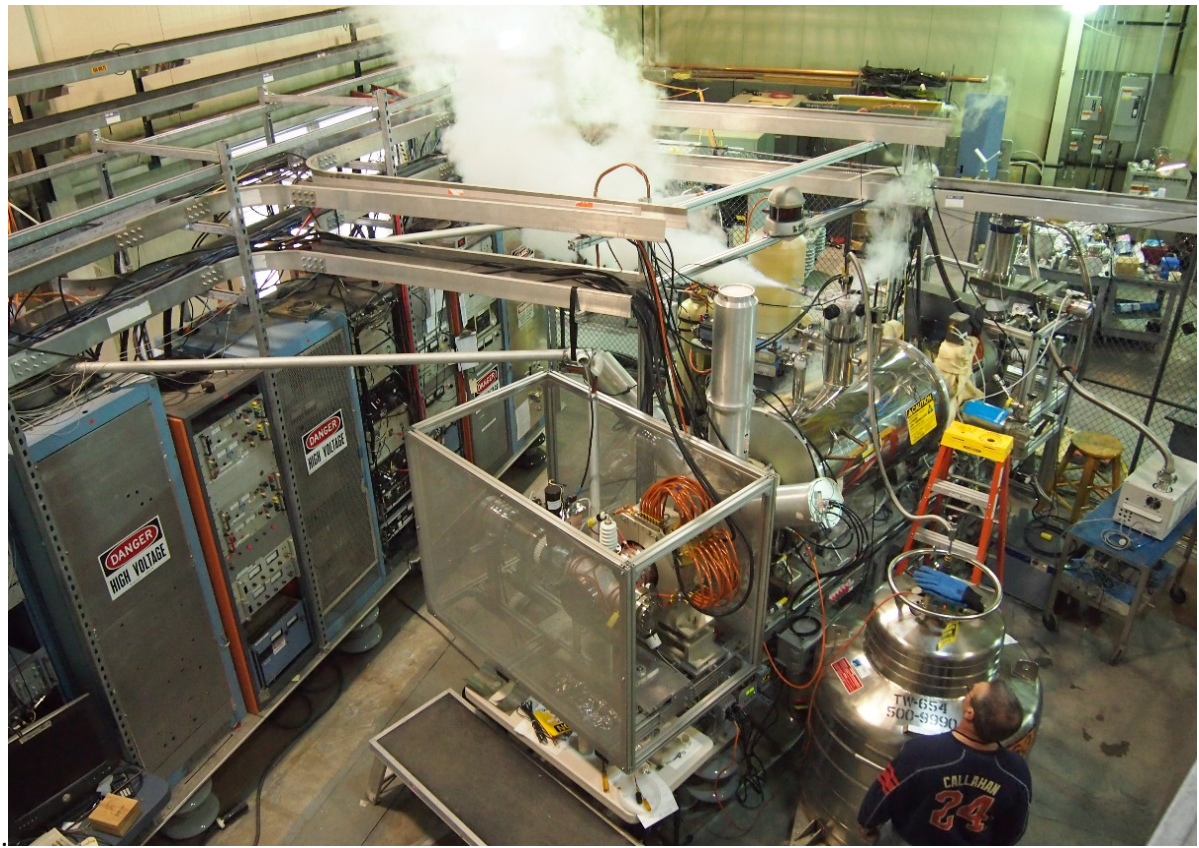

**Figure 15. The HEC² electron gun installed at TestEBIS.**

In the first run the beam was produced in short pulses with low duty cycle. The reason for short pulses was a power supply using a capacitor bank. Drawing 1.5 A current over 10 ms depleted the charge in the capacitors causing a sag of voltage and overall optics disturbance. Therefore the beam was extracted with a time-profile of the current as shown in Figure 17.

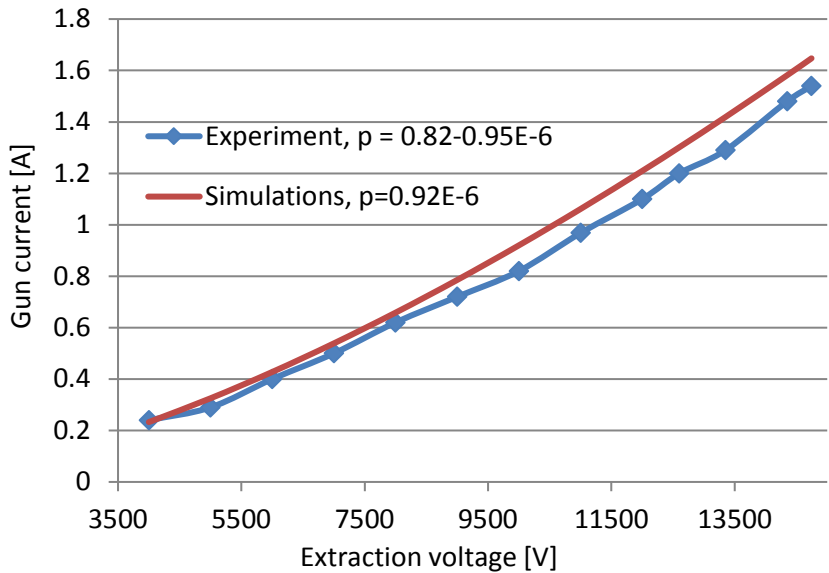

**Figure 16. Current extracted from the HEC² gun as a function of applied anode voltage.**

The duty cycle was limited by the legacy collector, designed for immersed beams. Immersed and Brillouin-like beams have mismatched energy deposition profiles [55]. Therefore, the collector cooling system designed for an immersed gun will be inefficient for Brillouin-like beams, and high-duty operation will cause local overheating and damage to the internal collector structure.

It is important to stress that the gun itself can do better and that limitations on the pulse length and duty cycle are imposed by the test stand.

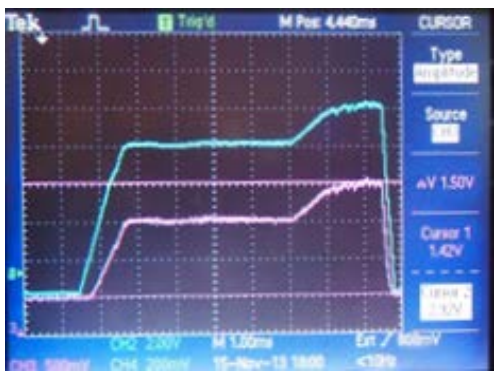

(A)

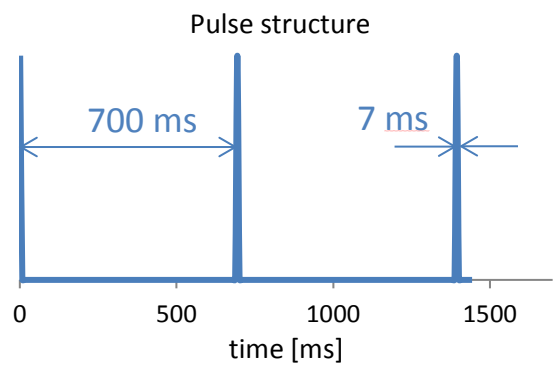

(B)

**Figure 17. Time structure of the gun pulses during the first experimental series. (A) An individual pulse of 7 ms, 1 A current extracted over 5 ms and thereafter increased to 1.5 A over 2 ms. The step is due to limits of the capacitor bank. (B) Duty cycle structure of the pulsed beam.**

The achieved electron current of 1.5 A was limited by a 20 mA loss current tripping the anode power supply [54]. Several possible sources for this current were analyzed [51], [55], [54]. The limiting loss current seems to be due to electrons extracted from the side surface of the cathode [54]. Such electrons with excessive transverse momentum will experience a magnetic mirror effect and reflection at the entrance of the high magnetic field region. An upgrade of the electron gun to mitigate side emission and introduce additional control over the gun geometry was tested later at BNL and allowed to achieve zero loss current at similar settings. During the second experimental run in spring 2014 a series of experiments were performed. The current was ramped to 1.7 A at about 35 keV electron energy in the ionization region. Given the applied magnetic compression and the size of the anode aperture, a lower limit of the current density could be set as ~ 200 A/cm², however, we expect the current density to be significantly higher than this. This expectation is based on numeric simulations of the optics and extraction of light, highly charged ions.

In preparation for the experiments TestEBIS was equipped with a Reflection Time-of-Flight Mass Spectrometer (RToFMS). In the absence of a primary ion injection beam-line, the extracted ions were created by ionization of the residual gas. Residual gas was ionized using a 450 mA electron beam with 16.6 and 8.8 ms pulse lengths. The ions were extracted from the trap and their charge state distribution (CSD) was measured (see Figure 18).

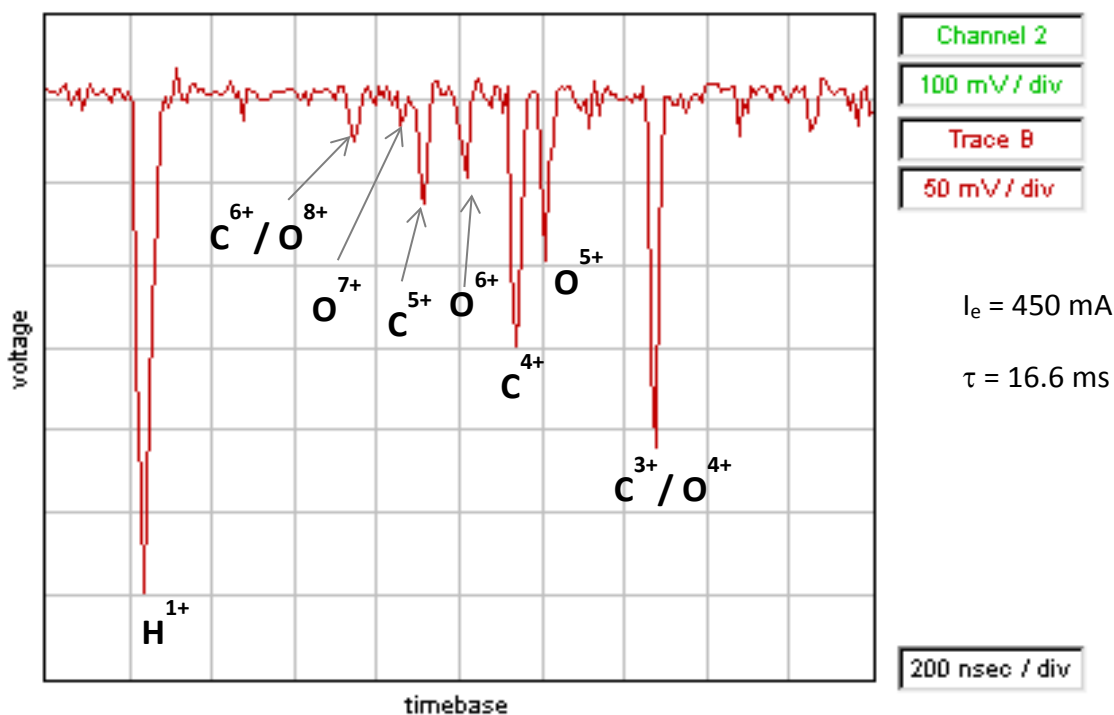

**Figure 18. Time-of-flight spectrum of extracted ions for a 450 mA electron current after 16.6 ms breeding time.**

In general, CSD can be used as a reliable instrument to measure the current density. In this preliminary experimental arrangement, however, we did not know the exact composition of the neutral atoms in the ionization region, which complicated the calculation. Estimations using ratios of $O^{6+}/O^{5+}$ and $C^{5+}/C^{4+}$ from the different spectra give results with too high error bars to be of practical use.

A side experiment was carried out to identify the presence of magnetron discharges in the electron gun chamber. The experiments reported by Pikin et al. in ref. [54] demonstrated that already at a few kV voltage between the anode (equipotential with the magnetic shield) and the gun chamber a deterioration of the vacuum was detectable. Suppressing the discharges by biasing the chamber can be an effective way to maintain better vacuum in the gun chamber and consequently the ionization region [54].

## 5.2   Subsequent tests and development

As a continuation of the initial commissioning tests, experiments were carried out with a modified gun. These modifications included isolating the focusing Wehnelt electrode and applying a negative bias voltage to reduce the extraction of electrons from the side cathode surface. Also two different mechanical versions of the gun were tried. One version included a movable cathode arm driven by a manual leadscrew mechanism. This design is shown in Figure 19.

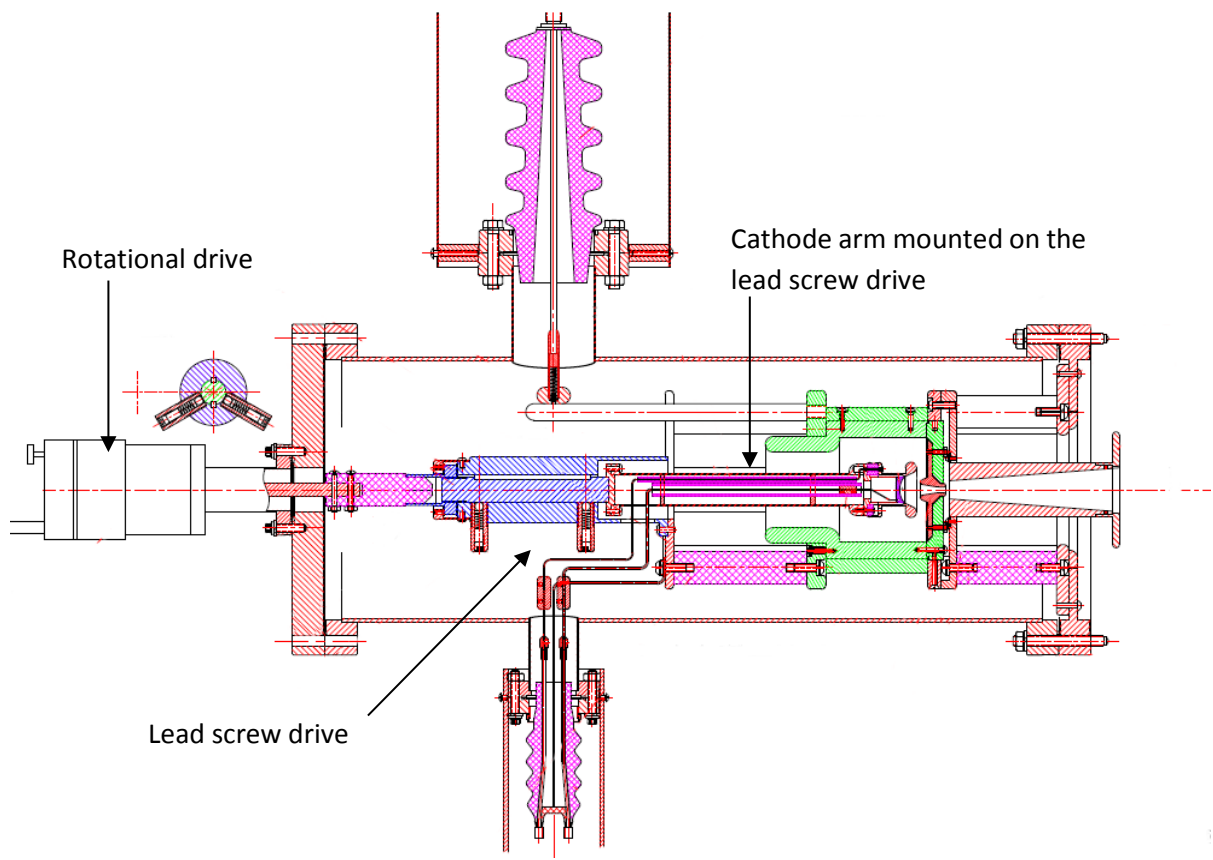

**Figure 19. The HEC² gun mechanical design with leadscrew translational mechanism (blue hatched area) driving the cathode arm.**

The other version included a fixed mounting on the back flange, allowing to mount and unmount the gun faster without removing the complete gun chamber and without using a 3-flange sandwich arrangement with the gun fixed to the double-sided alignment flange in the middle. The updated mechanical design reduced the number of mechanical joints and helped to maintain good radial alignment. This mechanical concept is shown in Figure 20.

The latter concept tested in June 2016 allowed to achieve a maximum current of 3.14 A, exceeding the present HEC² ISOLDE specification on current requesting values above 2.7 A. The compression was not measured due to lack of available methods. If the beam was compressed according to theory and simulations using the experimental conditions of a reduced magnetic field (3.3 T instead of 5 T), a current density of up to 6.9 kA/cm² was produced.

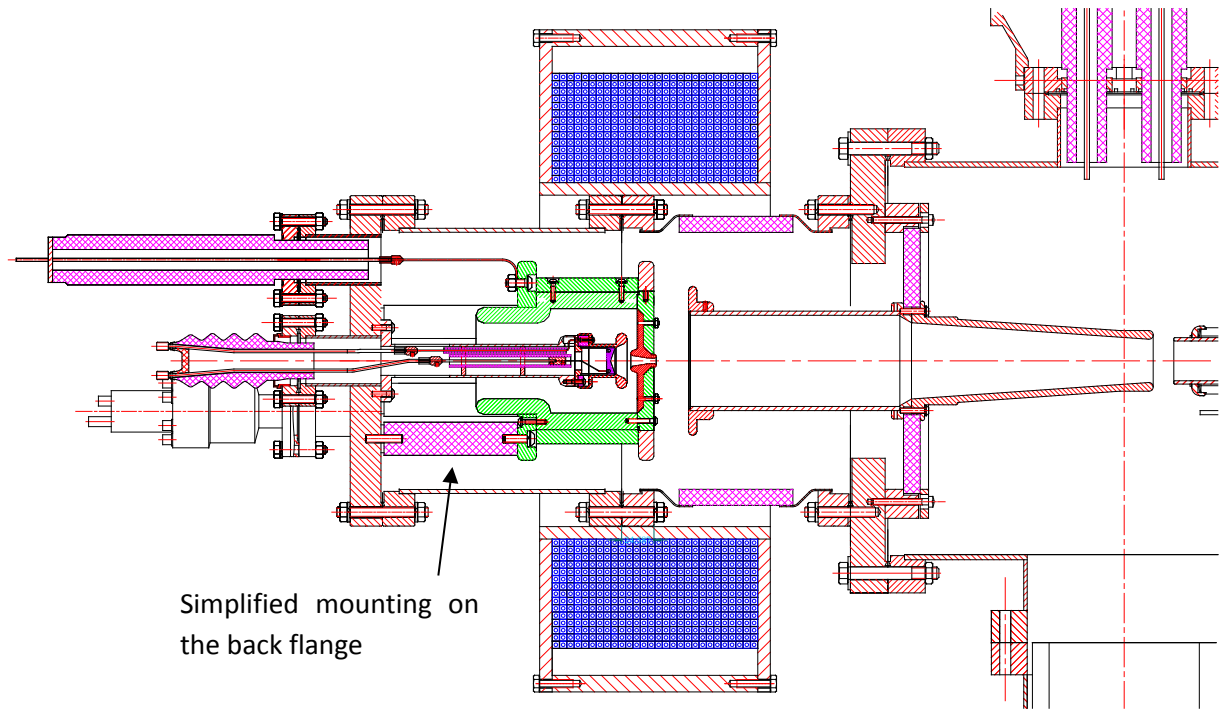

Simplified mounting on
the back flange

**Figure 20. HEC² gun mechanical design with simplified mounting on the back flange that was used in the final tests.**

It is important to mention that for currents below 2 A the beam transport was loss-free. The maximum extracted current of 3.14 A was limited by the available bias voltage, and at that point the anode current reached up to 23 mA, see Figure 21.

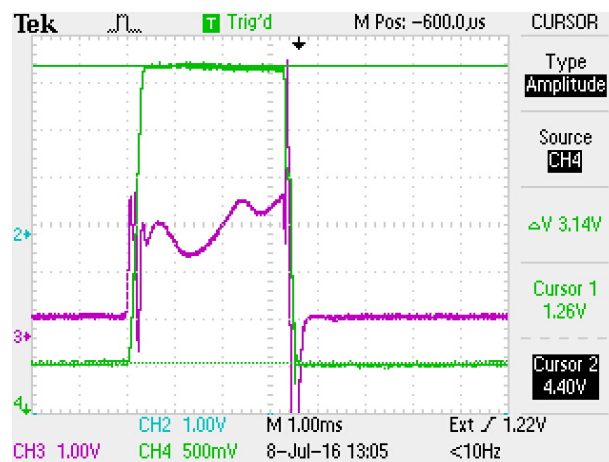

**Figure 21. Extracted electron current of 3.14 A (green) and anode current up to 23 mA (purple).**

## 5.3   *Beam diagnostics*

As a part of the design study two diagnostics devices were built to analyze the beam performance: a Pepper-Pot ion Emittance Meter (PPiEM) and a combined device for RTofMS and Transmission Energy Analysis (TEA). The RTofMS/TEA unit was shipped to BNL but was not yet used in operation (ToF mass spectra taken with MS borrowed from RHIC EBIS). See Figure 22 during the assembly stage. The PPiEM was built and commissioned at CERN and successfully used for first direct measurements of the REXEBIS emittance at ISOLDE [56].

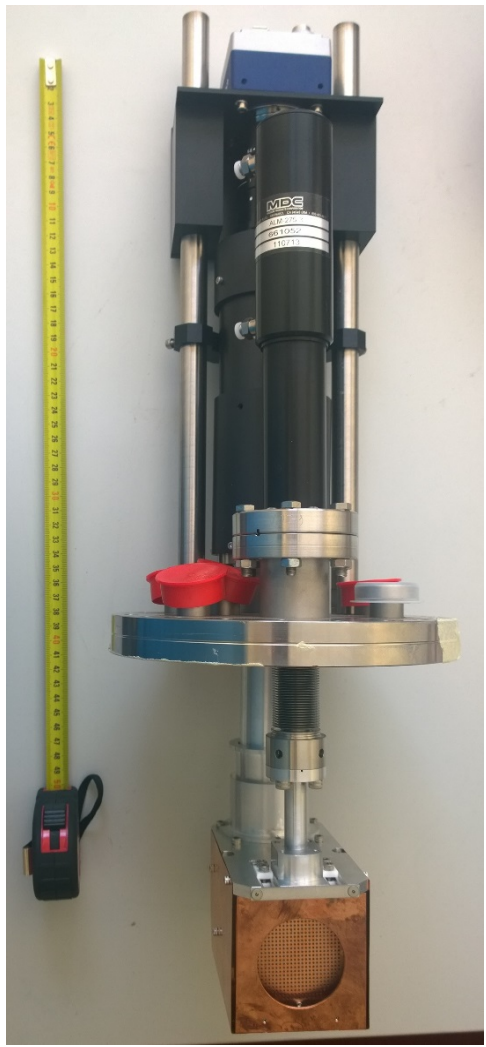
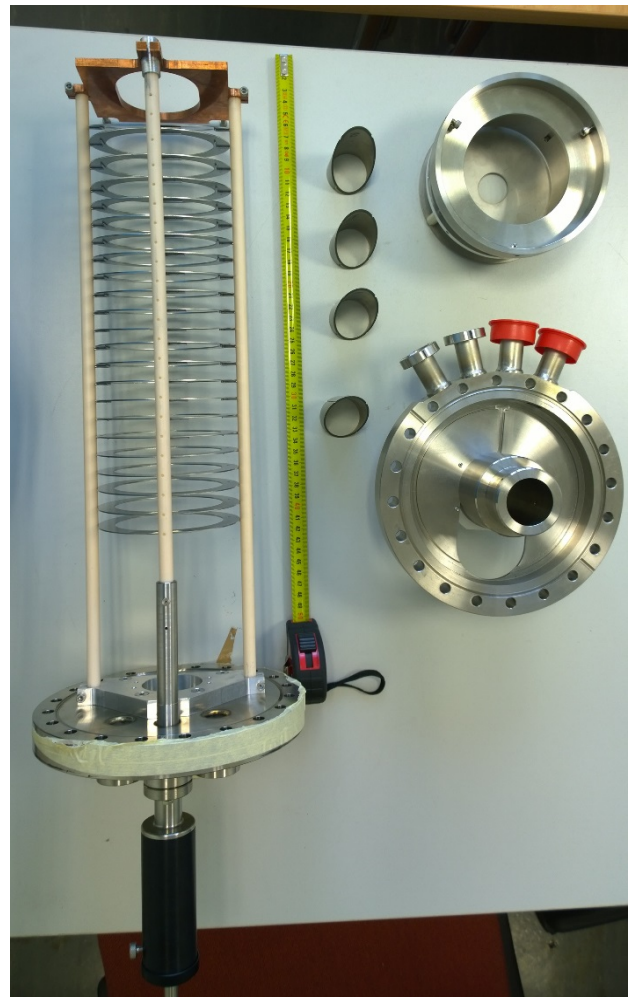

(A)                                    (B)

**Figure 22. Diagnostic equipment in production at CERN. (A) The ion emittance meter. (B) The reflection time-of-flight mass spectrometer combined with energy analyzer.**

## 5.4    Studied and refuted options

Several EBIT-like devices were influenced by the LLNL Super-EBIT. This family of devices consists of EBITs at MPIK, Tokyo University of Electric communications, University of Fudan (Shanghai), TRIUMF and MSU-NSCL. While designs keeping the original scale were performing close to specification (MPIK, TEC, UF), scaled-up versions (TRIUMF, MSU-NSCL) achieved only limited success in generating higher current while maintaining the current density. A substantial redesign effort now performed by MSU-NSCL may yield a more suitable high-current version of the LLNL-super-EBIT gun [42]. Before further results are obtained, pursuing up-scaling of the LLNL gun design should be considered as risky.

Most of the EBITs as well as several EBIS and all ESIS are built with a cryogenic internal volume, cold drift-tubes and in some cases even common vacuum of the magnet cryostat and the electron beam region. The latter option was used several times in the history of EBIS devices, for instance for CRYEBIS 2 (Saclay), KSU-CRYEBIS (Kansas State University), Frankfurt-CRYEBIS, Krion 1,2,3 EBIS (Dubna), CRYSIS (Stockholm), Super EBIS (Saturne), Dione (Saclay) [57], as well as for MSU-NSCL EBIT/S and many EBITs.

Other EBISes using cryo-pumping of the internal volume had a separation of the vacuum volumes like CRYEBIS 1 and Cornell EBIS 2. In general all these EBISes were demonstrating moderate vacuum performance in the $10^{-9}$ mbar region despite using some advanced cryogenics techniques like 2 K panels of super-cooled helium in Dione [57]. At the moment the use of common vacuum cryostat is strongly discouraged. Groups report of cold helium leaks from the LHe tank to the cryostat vacuum for warm bore magnets (BNL, ANL). At CRYSIS SE a substantial improvement of the vacuum conditions occurred [58] when the originally common vacuum was separated into a beam volume and magnet cryostat. The cryogenic option as such is also technologically not necessary. First demonstrated by Gobin at Saclay with a completely warm EBIS [59], it was later confirmed by REXEBIS, RHIC EBIS and CARIBU, with REXEBIS demonstrating unparalleled vacuum performance amongst all EBISes. This is partly due to the REXEBIS operating at lower current and without high voltage in the high magnetic-field region. It means that with a pumping scheme similar to the ANL CARIBU ECB, REXEBIS does not create conditions suitable for discharges causing non-thermal outgassing. For the future ECB it is important to reduce non-thermal outgassing caused by discharges, electron stimulated desorption and influx of residual gas from lower vacuum regions. A cryogenic scheme remains an option, but given its numerous drawbacks such as complex mechanics, reduced control over mechanical alignment, sensitivity to minor beam losses or energy transfer to cold surfaces in the order of few Watts (out of hundreds of kilowatt of the beam power) and an overall high maintenance, we would not recommend a cryogenics-based design.

### 5.5   Status of HEC²

As of July 2016 the joint research carried out by the CERN–BNL collaboration at BNL is concluded. The test facility used for experiments is being rebuilt for a polarized He source. In parallel CERN has re-commissioned its own EBIS test stand, TwinEBIS [60], which is an offline replica of REXEBIS. The HEC² gun however cannot be directly used at TwinEBIS as its mechanical design is not compatible. Therefore HEC² serves as a proof of the concept and an inspiration source for a new gun being built especially to fit TwinEBIS. Conclusions drawn from the HEC² test program influenced the design of the new gun and they are laid out in the following paragraph.

### 5.6   Development conclusion

Based on the performed tests several important conclusions were drawn for the future development. It was found that in general a multi-ampere Brillouin flow electron gun in the energy range of interest is feasible. The realization of the mechanical design is crucial for successful operation as the Brillouin optics is sensitive to misalignments. The precision alignment can be realized either by a system of sophisticated outside-vacuum adjustment plates, like in some Brillouin guns in the 80-s [17], or by a more integrated design leaving less room for misalignments to add up. The adjustment plates were reported to have poor reproducibility of settings [17]. Therefore an integrated design with rigid but well verified alignment seems as a safer choice. The positive influence of reduced mechanic complexity can be seen when comparing three tested configurations of the HEC² gun. For the future development it is important to keep in mind the risk of various discharges in the gun chamber. It would also be beneficial to design the future system such that the influence of thermal expansion on the gun geometry are mitigated, and that the gun is able to tolerate a substantial anode current. These objectives can be attained by adequate heat transfer between parts combined with water cooling.

## 5.7   Near future development opportunities

The experimental studies on a high performance ECB can be continued as a side topic of the presently ongoing MEDeGUN [40] program aiming to demonstrate a high repetition rate source of $C^{6+}$ ions for cancer radiotherapy with linac-based accelerators. MEDeGUN (Figure 23) is also a Brillouin type electron gun with an optics quite similar to HEC[2], but having the mechanical design highly integrated based on lessons learned with HEC[2]. The gun design puts strong emphasis on a high precision of the relative alignment of the critical elements, reduced number of mechanical joints, and reduced effects of thermal expansion owing to matching expansion coefficients and water cooling.

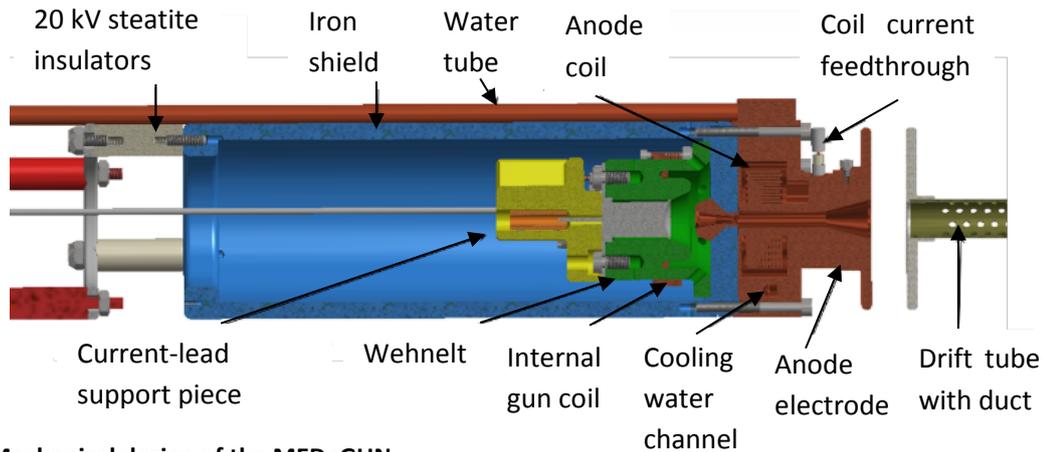

**Figure 23. Mechanical design of the MEDeGUN.**

MEDeGUN is being developed for an energy range below 10 keV so its application at higher energies will in fact reduce any space-charge related and beam optics issues. The target parameters of MEDeGUN are similar to the requirements of an upgraded ECB. The experimental studies will be performed using the TwinEBIS test bench, which was recently re-commissioned at the new location, see Figure 24. Using some of the beam diagnostics built for the HEC[2] tests will help us to perform comprehensive studies on the performance of the device.

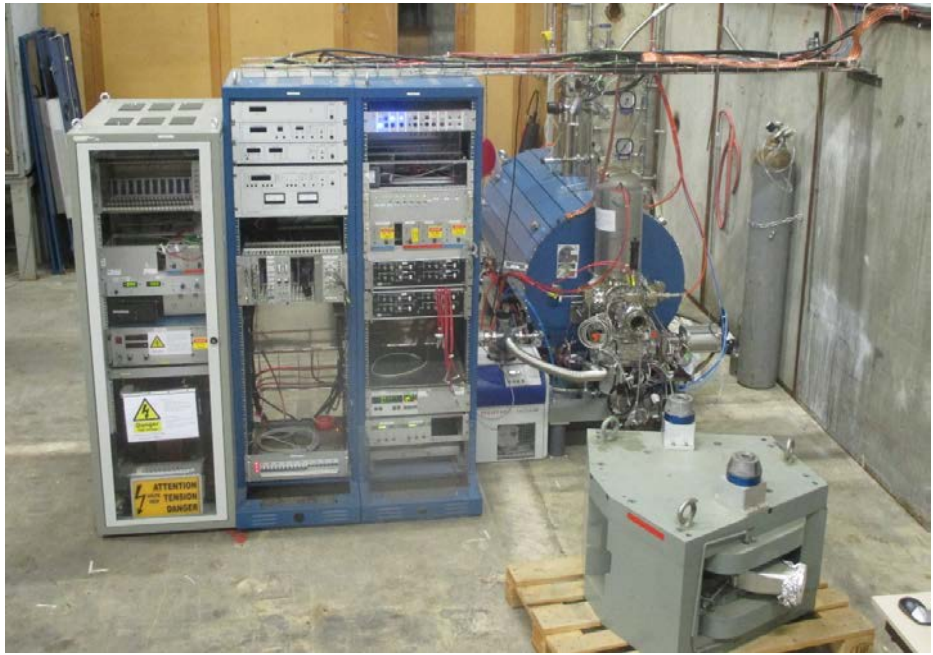

**Figure 24. TwinEBIS test bench re-commissioned at the new location in building 150.**

# 6   Cost estimate

In this section we give a financial overview of the project. As the project is still in the R&D phase the figures may evolve. Nevertheless, we can use the BNL RHIC pre-injector (including the high performance RHIC-EBIS) as guidance. The weighted contingency budget of the RHIC project was 24% of the total cost of material and labor [61]. It is consistent with DoE recommendations, where the contingency budget on a CDR phase project ranges from 25% (standard conditions) and 40% (experimental/special conditions) [62]. The latter is being the case for the HEC² project. As our reference projects are valued in USD while the budget is planned in CHF we introduce additional contingency to account for that. As the original analysis was based on 2014 exchange and inflation adjustment rates we have revisited it in 2016 to confirm that significant changes in the rates do not exceed our financial contingency.

## 6.1   Overview of spending so far

Cumulative costs of the HEC² program, not including manpower, covered by CATHI ITN amounted to 196 kCHF, with the cost breakdown structure given in Figure 25. The costs are dominated by hardware, production of the gun itself and beam diagnostics tools (to be transferred to further experiments). This includes the electron gun and its chamber, some complementary equipment, such as vacuum pumps and gauges, emittance meter, RToFMS/TEA, as well as shipping of the material to BNL, and sending CERN personnel.

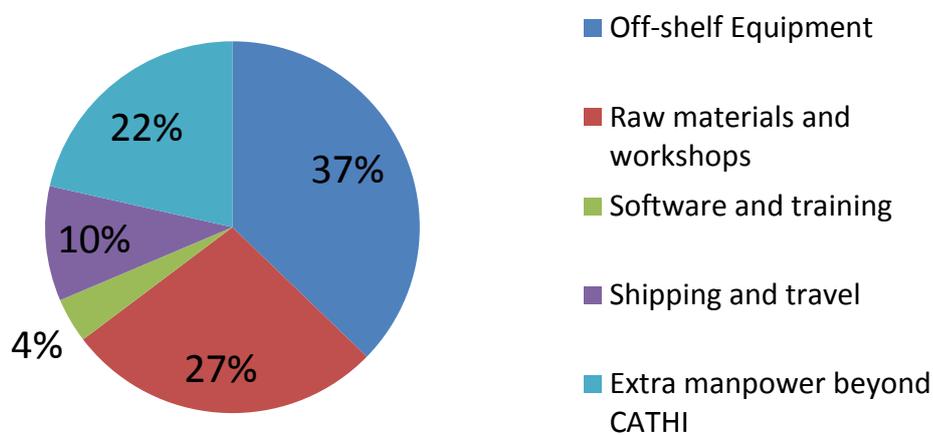

**Figure 25 . Cost breakdown for the HEC² project.**

## 6.2   Price estimate to build

In this section we estimate the cost required to construct the full ECB. In this estimation we use our own experience in building several sub-systems, the financial plans for RHIC-EBIS [61] and HICB [63] by MSU-NSCL. The total construction price is estimated to be 5573.1 kCHF. The breakdown of the costs is given in Table 2. Used estimation methods are labeled from A to K and are described in detail after the table.

**Table 2. Price estimation of HEC$^2$ compared to HICB and RHIC EBIS (2014 analysis)**

| System | CERN ECB | HICB | RHIC-EBIS |
|---|---|---|---|
| SC magnet | 450 kCHF (A) | 602.4 kCHF | 642 kCHF |
| EBIS structure | 363 kCHF (B) | 308.5 kCHF | 356 kCH |
| Vacuum system | 166 kCHF (C) | NA | 276.4 kCHF |
| Power supplies | 668.2 kCHF (B+D) | NA | 851 kCHF |
| Control system | 60 kCHF (C) | NA | NA |
| Beam diagnostics | 60 kCHF (C) | NA | NA |
| **Subtotal for main parts** | **1767.2 kCHF** | **1728 kCHF** (E) | **2125 kCHF** |
| Conversion uncertainty | 22% on A and B (F) | 6% (F) | 15% (F) |
| Uncertainty contribution | 324.7 kCHF | 103.7 kCHF | 318.8 kCHF |
| **Subtotal with correction** | **2091.9 kCHF** | **1831.7 kCHF** | **2443.8 kCHF** |
| Ratio of labor-to-material costs | 75% (G) | 77% (G) | 73% (G) |
| Cost of labor contribution | 1568.9 kCHF | 1410.4 kCHF | 1784 kCHF |
| **Subtotal with labor costs** | **3660.8 kCHF** | **3242 kCHF** | **4227.7 kCHF** |
| CERN-specific civil engineering | 320 kCHF (H) | Not applicable | Not applicable |
| **Subtotal of all** | **3980.8 kCHF** | **3242 kCHF** | **4227.7 kCHF** |
| Recommended contingency level [62] | 40% (K) | 40% (K) | 20-25% (K) |
| Contingency contribution | 1592.3 kCHF | 1296.8 kCHF | 972.4 kCHF |
| **Total** | **5573.1 kCHF** | **4539 kCHF** | **5200 kCHF** |

***Estimation and conversion methods:***

A: Price scaling law by Green and Strauss [64] for the solenoid magnets was used with a fitting parameter taking into account accuracy requirements of EBIS magnets. The scaling was introduced in the way that the scaled equation will give the price of the RHIC EBIS magnet, when its geometric parameters are substituted. The required scaling factor is 1.2.

B: Taking the item list and prices from RHIC EBIS, adjusted for inflation (see F) and converted to CHF.

C: Item list based on RHIC EBIS, prices according to CERN catalog.

D: Direct quotes by suppliers.

E: From the 3784 k$ total price we excluded 20% overhead according to [63], divided by 1.77 for 0.77 labor-to-material price ratio for MSU (see paragraph G). The obtained USD (2010 FY) price is then inflation adjusted and converted to CHF (see paragraph F).

F: When using comparison to RHIC-EBIS (HICB) the prices are converted from 2005 (2010) US dollars to present CHF by using US inflation adjustment factor of 1.21 (1.09) and currency exchange rate for USD to CHF conversion of 0.89. The financial error bar of the price conversion can be estimated as the difference between coefficients of scaling by USD inflation adjustment and USD to CHF conversion with 2014 exchange rate, and exchange USD to CHF using historical rates and then CHF inflation adjustment. Historical exchange rates and inflation adjustment factors taken on 01.09.2010. The first method gives 1.21·0.89=1.077 (1.09·0.89=0.97) conversion for 2005 (2010). The second method 1.038·1.265=1.313 (0.9895·1.05=1.03). The additional error bar due to currency conversion is 22% (6%) for 2005 (2010) USD prices. This error bar is included as currency conversion uncertainty in the table.

G: The analysis of the detailed WBS for both proposals [63] and [61] shows that the cost of labor is about 70-80% of the material cost reaching 73% for all WBS of the RHIC-EBIS project combined (sum of all labor costs and all the material) and 77% for MSU contribution to HICB project (major contribution and the only one given with comprehensive breakdown structure). As both proposals were calculated based on US labor cost and efficiency this should be corrected while applying to CERN. We assume 75% labor cost contribution for the CERN project.

H: Additional CERN-specific civil engineering and integration into ISOLDE including the cost of labor. This includes additionally rebuilding of the HV cage and shielding, cabling and distribution of 200 kW electric power in the ISOLDE hall, and the installation of 100 kW demineralized cooling water capacity.

Both sufficient electric power and water cooling capacities have been confirmed (R. Necca and P. Pepinster). Electric price based on EMTE quote and required material.

K: The contingency budget accounts for unforeseen costs driven by both technological risks (unforeseen material cost) and scheduling risks (additional costs of manpower). In this calculation, we use recommendations by the US Department of Energy (DoE) [62]. The recommended value is a percentage of the total cost, where the percentage depends on the stage of the project (CDR or TDR), its nature (civil engineering or research equipment) and level of technical risk. In our case we assume special/experimental conditions and a CDR stage of the project. In this case 40% contingency budget is recommended. For the BNL project, the authors used values of 20-25% depending on the work package, and we use 23% in the table. The lower contingency budget is motivated by their full-scale tests at TestEBIS and the more expensive preliminary R&D program.

### 6.3    *Revisiting the budget (2016)*

Over 2014-2016 exchange rates of CHF, EUR and USD have substantially changed. Using the new exchange rates and inflation adjustment (for USD and for CHF) we have recalculated Table 2 regarding HIE-EBIS costs. Using CHCPI2011 and USCPI31011913 indexes we obtain an inflation adjustment for USD 2005-2016 of 1.23 and 0.99 for CHF, while the USD to CHF exchange rate decreases from 1.265 to 0.965 between 1 Sep 2005 and 2016. As such, when comparing analyses from 2014 and 2016 we obtain a higher evaluation of the prices converted from USD to CHF (due to extra inflation adjustment and higher appreciation of USD), while we see a smaller financial uncertainty (K) as 1.23·0.965=1.187 and 1.265·1.03=1.253 giving only 9.8% exchange rate uncertainty.

We have additionally recalculated the costs of the main magnet as the specification of the electron optics was updated. The specification now features a lower electron current, which in turn calls for a longer trapping region to maintain the trapping capacity. Also, in order to counteract TSI a higher magnetic field is required. The new magnet requirements look as follows (old values in brackets): main field 6 T (5 T) and trapping region length 1 m (0.7) m. The requirements on the stray field remain the same. We recalculate the magnet price in the following way (L in Table 3). The best offer for magnet fulfilling the old specification (by Cryomagnetics Inc) is fitted by introducing a scaling coefficient to Green and Strauss formula inflation-adjusted to 2016 USD. Then we substitute 6 T field and 1.3 m winding length into the corrected formula. Currency exchange uncertainty of 9.8% for 2016 to be applied as original pricing is in USD. This analysis also points to the fact that the price estimate for the magnet in the original 2014 cost estimate was too optimistic. Using the same approach to the original magnet specifications we arrive to 525 kCHF, i.e. substantially higher than the old estimation of 450 kCHF.

The cost breakdown according to the new specifications and the financial market situation is shown in Table 3. The change on the financial markets alone caused the cost estimate to change by -0.8%. In the same time new magnet specifications and more accurate quote-based cost estimation had driven equipment and related costs to a total of 6109.5 kCHF, i.e. 9.6% above the original estimate.

It is important to note, that in case of successful tests of MEDeGUN in heavy-duty operation, one could reasonably cut the contingency budget to 20-25%, a value similar to the RHIC-EBIS case if the fully-functioning similarly-scaled prototype is built and tested.

**Table 3 . Revision of the cost estimate due to changes on financial markets in 2014-2016.**

| System | CERN ECB (2014) | CERN ECB (2016) |
|---|---|---|
| SC magnet | 450 kCHF (A) | 703.3 kCHF (L) |
| EBIS structure | 363 kCHF (B) | 396.8 kCHF (B) |
| Vacuum system | 166 kCHF (C) | 173.3 kCHF (C) |
| Power supplies | 668.2 kCHF (B+D) | 736.6 kCHF (B+D) |
| Control system | 60 kCHF (C) | 60.0 kCHF (C) |
| Beam diagnostics | 60 kCHF (C) | 60.0 kCHF (C) |
| **Subtotal for main parts** | **1767.2** kCHF | **2130.0** kCHF |
| Conversion uncertainty | 22% on A and B (F) | 9.8% on A, B, L (F) |
| Uncertainty contribution | 324.7 kCHF | 180.8 kCHF |
| **Subtotal with correction** | **2091.9 kCHF** | **2310.8 kCHF** |
| Ratio of labor-to-material costs | 75% (G) | 75% (G) |
| Cost of labor contribution | 1568.9 kCHF | 1733.1 kCHF |
| **Subtotal with labor costs** | **3660.8 kCHF** | **4043.9 kCHF** |
| CERN-specific civil engineering | 320 kCHF (H) | 320 kCHF (H) |
| **Subtotal of all** | **3980.8 kCHF** | **4363.9 kCHF** |
| Recommended contingency level [62] | 40% (K) | 40% (K) |
| Contingency contribution | 1592.3 kCHF | 1745.6 kCHF |
| | | |
| **Total** | **5573.1 kCHF** | **6109.5 kCHF** |

# 7 Collaboration partners / programs

The HEC[2] project was carried out in collaboration with BNL taking advantage of available infrastructure and expertise. As of June 2016 the project is concluded on the BNL side, the BNL TestEBIS is being rebuilt into a polarized He source. Some studies are possible using the recently recommissioned test bench TwinEBIS [60], for example the commissioning of MEDeGUN [40].

## 7.1 ENSAR-2 for EURISOL program

An ECB similar to the studied one is of great interest for a future EURISOL facility. Fast breeding will improve the overall performance of EURISOL [65]. A higher throughput capacity is of limited use for ISOLDE, but may be requested by EURISOL. The ENSAR-2 started in April 2016 and partly funds manpower to perform studies on the EURISOL ECB.

### *7.2    TSR@ISOLDE as the most demanding EBIS user worldwide*

The TSR@ISOLDE experimental program puts the highest constraints on the ECB performance as intense ion beams of extremely high charge states are requested. None of the other projects is aiming at such exceptional performance. Some of the experiments at TSR@ISOLDE will not just benefit from a high performance ECB, but in fact require it for their realization.

### *7.3    Benefits for other CERN projects*

Several projects at CERN may benefit indirectly from the HEC$^2$ ECB project. If successful, HEC$^2$ ECB may be a proof-of-principle device for the once proposed but dismissed LHC-EBIS for heavy ion injection to the LHC [66]. Another possible beneficiary is the HL-LHC project. To suppress the beam halo it was suggested to install electron lenses in the LHC [67]. Electron lenses were successfully used at Tevatron and are now under commission at RHIC [68]. The LHC electron lens requires 63 Am [69], i.e. at a practically feasible length of a few meters it requires a multi-ampere electron beam. An electron lens has many similarities with an EBIS in the beam optics. Moreover, in the case of BNL, the electron lens project relied heavily on the EBIS team expertise and used for testing purposes exactly the same TestEBIS as we are now using for the HEC$^2$ experiments [68]. Therefore, in-house expertise in handling multi-ampere beams, performing reliable simulations of them, manufacturing technologies, know-how and other results gained in the HEC$^2$ ECB project can bring a long-term benefit to CERN.

### *7.4    Benefits outside the nuclear and particle physics community*

Development of an EBIS with both high pulse intensity and repetition rate has potential benefits beyond the nuclear and particle physics community. From mid-90s EBISes have been proposed [70] and studied [39] as ion sources for ion cancer therapy. ESISes working in ISOL mode were studied as a source of $^{11}C^{6+}$ nuclei, a positron emitting carbon isotope allowing to combine ion therapy with in-situ dose verification by means of positron emission tomography [71]. EBISes were studied [72] for second generation ion treatment facilities such as cyclinac [73] and ion Rapid Cycling Medical Synchrotrons (iRCMS) [74] where the delivered ion energy is adjusted by the accelerator settings and is switched between fast cycling pulses. Contemporary EBISes with moderate current density achieve only repetition rates suitable for iRCMS, that is around 30 Hz [32], and are under further study at the Heidelberg Ion Therapy center in Heidelberg, Germany. Several other types, such as cyclinac, high frequency linac [75] and Fixed Field Alternating Gradient (FFAG) [76] designs demand repetition rates of 100-400 Hz [77]. This requires a high-current-density EBIS [78]. To facilitate the development of second generation ion treatment facilities a spin-off of the HEC$^2$ project, MEDeGUN, was proposed to the CERN Knowledge Transfer Fund and was awarded funding in February 2015. As of spring 2016 the gun is in production with the installation planned in autumn 2016.

## 8    Conclusions

In this report we summarized the results of R&D carried out in 2012-16 on a high performance charge breeder for HIE-ISOLDE and related projects in the CATHI ITN framework. During the course of parametric studies it was found that the most technologically challenging part of the project is a high compression multi-ampere electron beam required to achieve the desired charge states within the specified time, while maintaining the pulse intensity and the acceptance for primary ions. Experiments with such a gun built within the design study and tested at the BNL TestEBIS have shown promising results with achieved current exceeding the design specifications. There is no solid number on the actual value of achieved beam compression and the only way to estimate it is to assume a good agreement with simulation. A poor compression would very likely manifest itself in a significant loss current. As the tests were performed with a legacy collector not optimized for the HEC$^2$ optics the heavy-duty cycle was not demonstrated and can be shown only with a fully dedicated beam optics. Such an opportunity is available in the framework of the MEDeGUN project, where heavy-duty operation of an EBIS with a high-compression beam is the goal.

Technological advances achieved on the way to the full design specification will serve not only the HIE-ISOLDE project, but several other areas, including particle physics (knowledge transfer to electron lenses used in HL-LHC), next generation ISOL-facilities (TSR@ISOLDE, EURISOL) and ion cancer therapy (MEDeGUN).

# 9    Acknowledgements

We would like to thank the following people who contributed to this project in different ways:

BNL team involved in experimental studies: J. Alessi, E. Beebe, R. Schoffer, D. McCafferty

CERN management: R, Catherall, Y. Kadi, R. Scrivens, A. Lombardi

CERN workshop teams: J.-M. Geisser, E. Barbero, E. Rigutto, P. Moyret, J. Thibout, M. van Stenis

CERN material scientists and specialists: S. Sgobba, A. Teixeira

CERN scientists from collaborating projects: D. Voulot, M. A. Fraser, I. Podadera-Aliseda, E. Siesling

MSU-NSCL colleagues who shared their expertise: S. Schwarz, A. Lapierre, K. Kittimanapun

Colleagues who shared their simulation codes: R. Becker (University of Frankfurt, CBSIM), Y. Zou (University of Fudan, CBSC), X. Lu (University of Electronic Science and Technology of China, Chengdu, NSCB).

Special thanks to R. Marrs (LLNL) for his input on ion heating, cooling and plasma instabilities.